\documentclass[preprint,12pt,authoryear]{elsarticle}

\usepackage{amssymb}
\usepackage{amsmath}

\usepackage{mhchem}
\usepackage{siunitx}
\usepackage{booktabs}
\usepackage{graphicx}
\usepackage{rotating}
\usepackage{xcolor} 

\usepackage{hyperref}

\usepackage{silence}
\ErrorsOff*

\usepackage[hyphens]{url}
\newcounter{verbctr}
\newenvironment{labelverb}{%
  \refstepcounter{verbctr}%
  \begin{tabular}{@{}p{0.8\textwidth}r@{}}%
}{%
  & \quad(\theverbctr) \\[1ex]%
  \end{tabular}%
}

\journal{Journal of Food Engineering}

\begin{document}

\begin{frontmatter}



\title{Visible and Hyperspectral Imaging for Quality Assessment of Milk: Property Characterisation and Identification} 


\author[a]{Massimo Martinelli\corref{cor1}\fnref{equal}}
\ead{massimo.martinelli@isti.cnr.it}
\author[b]{Elena Tomassi\fnref{equal}}
\author[b]{Nafiou Arouna}
\author[b]{Morena Gabriele}
\author[b]{Laryssa Perez Fabbri}
\author[b]{Luisa Pozzo}
\author[c]{Bianca Castiglioni}
\author[c]{Paola Cremonesi}
\author[d]{Giuseppe Conte}
\author[a]{Davide Moroni}
\author[b]{Laura Pucci}

\cortext[cor1]{Corresponding author}
\fntext[equal]{These authors contributed equally to this work}

\address[a]{Institute of Information Science and Technologies, National Research Council, via G. Moruzzi, 1, Pisa, 56124, Italy}
\address[b]{Institute of Agricultural Biology and Biotechnology, National Research Council, via G. Moruzzi, 1, Pisa, 56124, Italy}
\address[c]{Institute of Agricultural Biology and Biotechnology, National Research Council, via dell'Università 6, 26900 Lodi}
\address[d]{Department of Agricultural, Food, and Agri-Environmental Sciences, via del Borghetto, 80, 56124 Pisa, Italy}

\begin{abstract}
Rapid and non-destructive assessment of milk quality is crucial to ensuring both nutritional value and food safety. In this study, we investigated the potential of visible and hyperspectral imaging as cost-effective and quick-response alternatives to conventional chemical analyses for characterizing key properties of cow’s milk. A total of 52 milk samples were analysed to determine their biochemical composition (polyphenols, antioxidant capacity, and fatty acids) using spectrophotometer methods and standard gas–liquid (GLC). Concurrently, visible (RGB) images were captured using a standard smartphone, and hyperspectral data were acquired in the near-infrared range. A comprehensive analytical framework, including eleven different machine learning algorithms, was employed to correlate imaging features with biochemical measurements. Analysis of visible images accurately distinguished between fresh samples and those stored for 12 days (100\% accuracy) and achieved perfect discrimination between antibiotic-treated and untreated groups (100\% accuracy). Moreover, image-derived features enabled perfect prediction of the polyphenols content and the antioxidant capacity using an XGBoost model. Hyperspectral imaging further achieved classification accuracies exceeding 95\% for several individual fatty acids and 94.8\% for treatment groups using a Random Forest model. These findings demonstrate that both visible and hyperspectral imaging, when coupled with machine learning, are powerful, non-invasive tools for the rapid assessment of milk's chemical and nutritional profiles, highlighting the strong potential of imaging-based approaches for milk quality assessment. 
\end{abstract}

\begin{graphicalabstract}
\includegraphics[width=1.0\textwidth]{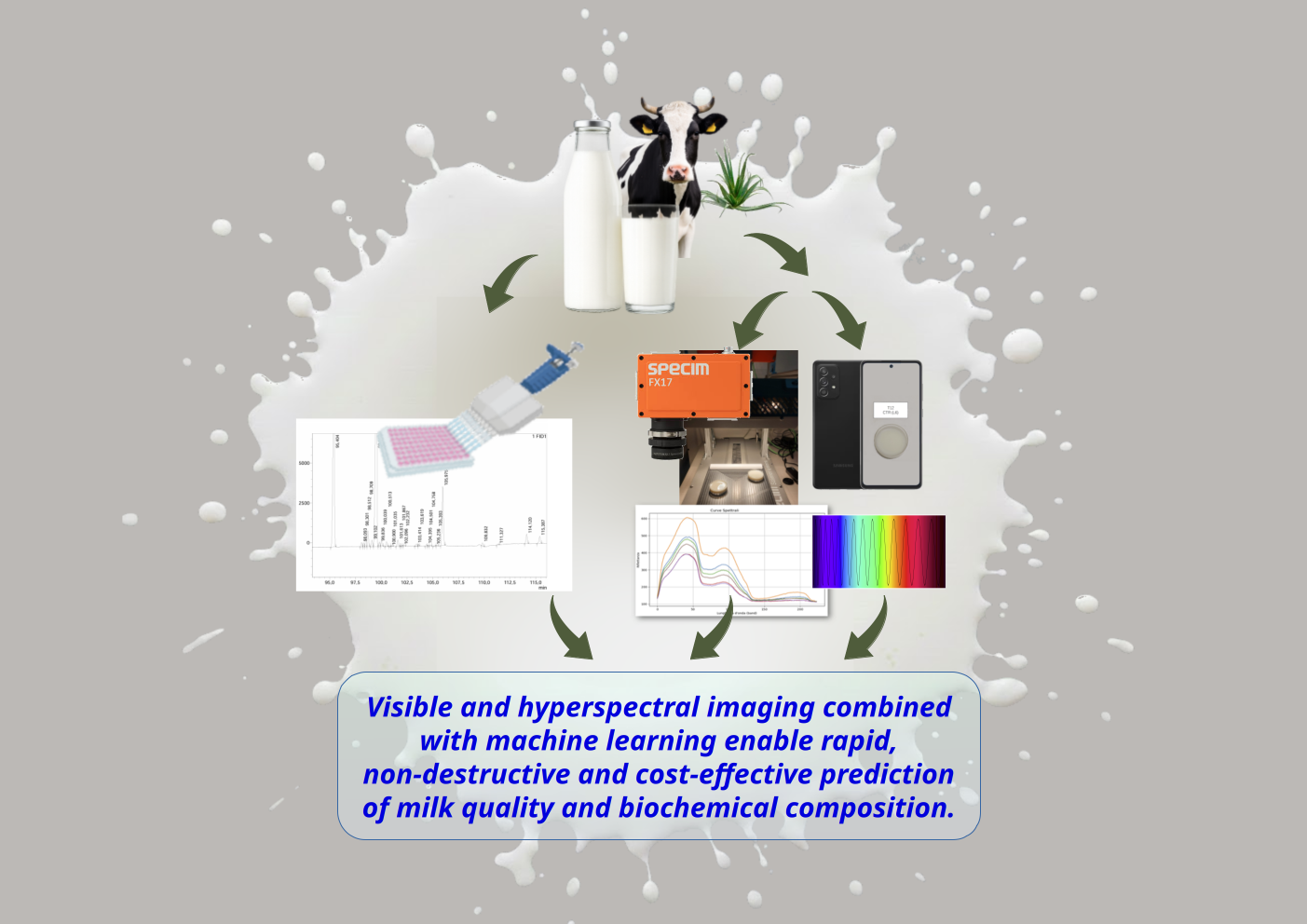}
\end{graphicalabstract}

\begin{highlights}
\item Visible and hyperspectral imaging enable rapid, non-destructive milk analysis
\item RGB imaging achieves 100\% accuracy in discriminating storage time and antibiotics
\item Image features predict polyphenols and antioxidant capacity with high precision
\item Hyperspectral imaging achieves over 95\% accuracy in fatty acid classification
\item Combining imaging and machine learning offers a cost-effective alternative to chemical analysis

\end{highlights}


\end{frontmatter}



\section{Introduction}
\label{sec1}

Cow's milk is a complete and nutritionally significant food, providing an important source of essential nutrients for the human body (\citet{khan2019antioxidant}). Milk and dairy products are rich in essential fatty acids, vitamins, minerals, and bioactive compounds, such as antioxidants, that contribute to both their nutritional value and oxidative stability. According to \cite{haug2007}, measuring the total antioxidant capacity of milk can provide valuable information about its biochemical status and nutritional value.
The fatty acid profile represents another essential determinant of milk quality. Milk fat contains a diverse set of fatty acids, including oleic acid, omega-3 fatty acids, and conjugated linoleic acid (CLA), whose concentrations are strongly affected by feeding practices (\citet{GOMEZCORTES20181}). Diets rich in fresh pasture, for example, can significantly increase the proportion of CLA and omega-3 fatty acids while lowering the omega-6/omega-3 ratio, an important nutritional marker associated with positive health outcomes (\citet{Kholif2022523536}). The fatty-acid composition is not only a nutritional indicator but also a potential fingerprint for authentication, traceability, and quality monitoring.
The variations in these components depend on several factors, such as the animal’s diet and management conditions, making them important indicators for evaluating milk’s functional and nutraceutical properties.
In addition to nutritional and dietary factors, milk quality is influenced by animal health and veterinary treatments. Milk obtained from cows undergoing antibiotic therapy can exhibit biochemical differences linked not only to potential drug residues but also to the underlying infection, typically mastitis, that prompted the treatment (\citet{ZHANG2022111969}). 
The problem of antibiotic resistance in veterinary and human medicine has become increasingly serious over the years (\citet{ZHANG2022111969}; \citet{microorganisms11041065}). This has increased the risk of ineffective treatment of bacterial infections (\citet{vanHoeij2018}). However, the use of antibiotics may increase bacterial resistance, posing a risk to the treatment of human and animal bacterial infections. 
Even when withdrawal periods prevent contamination with residues, physiological stress and inflammation may alter milk properties, influencing parameters such as oxidative status and fatty-acid distribution. By contrast, milk from untreated and clinically healthy cows generally shows more stable compositional profiles. Distinguishing these two categories is therefore relevant for both safety and quality evaluation, particularly when employing techniques sensitive to biochemical variations.
In recent years, nutraceuticals and bioactive feed additives, have shown promising effects in modulating rumen fermentation, shaping gut microbiota, and supporting immune and metabolic functions (\citet{10.3389/fvets.2025.1727126}). Their activity is largely linked to antioxidant, antimicrobial, and immunomodulatory properties. Among these natural products, Aloe species, such as \textit{Aloe vera}, have attracted considerable interest for their biologically active compounds, including vitamins, minerals, enzymes, polysaccharides, phenolics, and organic acids, associated with anti-inflammatory, immune-stimulating, antimicrobial, antioxidant, and lipid-modulating effects (\citet{Bani02042016}; \citet{Cremonesi2024}). 
Given the relevance of these compounds, analytical approaches capable of rapidly and non-destructively assessing milk composition are increasingly important. In this context, visible and hyperspectral imaging techniques offer powerful tools for monitoring quality attributes, including antioxidant activity, polyphenol content, and fatty acid profiles, providing a more comprehensive understanding of milk’s biochemical status.
Several studies have demonstrated that VIS–NIR spectroscopy can effectively predict antioxidant capacity, lipid oxidation, fatty-acid composition, and phenolic content in various food matrices (\citet{fodor2024role}; \citet{martinez2024multispectral}). In dairy research, spectral signatures have been shown to correlate with fat content, protein fractions, somatic cell count, and even specific fatty acids, making these methods promising for real-time, in-line assessment of milk quality (\citet{gastelum2020optical}).
Integrating hyperspectral data with milk’s known biochemical variability—such as antioxidant components, fatty acid patterns, and health-related alterations—offers an opportunity to develop advanced predictive models capable of rapidly and non-invasively evaluating milk quality attributes. This approach aligns with broader trends in precision livestock farming and digital dairy analytics, where spectroscopy-based tools are increasingly adopted to support quality control, authentication, and nutritional profiling.

Significant scientific attention has been focused on evaluating its antioxidant and anti-inflammatory properties, with studies mainly measuring polyphenols and fatty acids. 
The overall antioxidant capacity was quantified using the Ferric Reducing Antioxidant Power (FRAP) assay, a well-established method that reflects a sample’s potential to combat oxidative stress by measuring its ability to reduce ferric ions (Fe3+) to ferrous ions (Fe2+). Both polyphenols, natural compounds essential for neutralizing free radicals and influencing key cellular mechanisms, and fatty acids, lipid molecules vital for cell structure and energy metabolism, are key focus areas in this work.
More precisely, polyphenols are organic compounds characterized by multiple phenolic groups (aromatic rings with hydroxyl groups -OH), and fatty acids are biochemical compounds of molecules made up of long chains of carbon atoms that can be linked by single bonds, called saturated, or double bonds, called unsaturated.
In addition to its nutritional profile, a key area of research concerns milk adulteration and fraud. The dilution of milk with water is a widespread form of adulteration, and non-destructive methods are important to detect minute alterations in the absorption bands of water or other contaminants, such as urea, hydrogen peroxide, antibiotics, detergents, or acids, which are sometimes added to mask defects or alter the nutritional profile of the product.
Various analytical techniques are employed in the scientific field for the analysis of milk and liquid samples in general, among these, gas-liquid chromatography (GLC - \citet{james1956gas}), and hyperspectral imaging (HIS - \citet{goetz1985imaging}).
GLC is an analytical technique in which volatile compounds are separated based on their differential partitioning between a gaseous mobile phase and a liquid stationary phase coated on a solid support within a column. The sample is vaporized and carried by an inert gas through the column, where components separate according to their interactions with the stationary phase and are subsequently detected to generate a chromatogram for qualitative and quantitative analysis (\citet{albukhaiti2017gas}; \citet{imperiale2023analysis}).

Hyperspectral imaging (HSI) is an advanced analytical technique that integrates conventional imaging and spectroscopy to characterize materials by acquiring a three-dimensional dataset, called a hypercube, in which each pixel contains a continuous spectrum. Its fundamental principle is based on the quantification of the interaction between light and matter, and the resulting spectra, derived from reflected, transmitted, or emitted light, provide detailed information about the physicochemical properties of the sample being analysed.
Hyperspectral imaging extends the capabilities described above beyond the visible spectrum by acquiring measurements of electromagnetic radiation across numerous contiguous, narrow spectral bands, allowing it to capture detailed spectral signatures at each image point and provide comprehensive information about the composition and intrinsic properties of the observed materials.
These images can be processed and interpreted to extract meaningful information and enable automated actions through Computer Vision, a field of Artificial Intelligence that mimics human vision to derive insights. Subsequently, each type of extracted data can be analysed using traditional Machine Learning methods or advanced Deep Learning techniques, facilitating comprehensive data processing and supporting informed choices.
Many works in the literature explore hyperspectral data \citet{hebling2022} provides a systematic review of these works. In a number of studies, the Raman method is used: since it is generally considered a non-destructive and non-invasive technique, anyway some sensitive samples may be affected by laser radiation, such as heating or local changes: \citet{molecules28062770}, \citet{Yang2022}, \citet{jianwei2014}, \citet{qin2013998}, \citet{qin01022017},\citet{TAYLOR2025142035},\citet{Walter2015}. 
Some works with the Raman method were also carried out on powdered milk [\citet{s16040441}, \citet{Lee03062018}, \citet{s20164645}, \citet{LIM2016183}, \citet{Zhang2025}, \citet{Li2018}, \citet{YangNiu2022}]  to determine different adulterations.
Hyperspectral method has been also used in many cases: in \citet{Aqeel2025}, \citet{ZhangLiu2025}, the authors demonstrate its use to determine the adulteration of Salicylic acid, boric acid, glucose or formalin; in \citet{Unger2022} the presence of Escherichia coli or Listeria monocytogenes is determined in milk and in different cheeses is demonstrated; in  \citet{Sekhon2024} the efficacy is evaluated in determining the presence in milk and in milk powder of Listeria monocytogenes, four strains of Escherichia coli, six strains of the "Big Six" di Escherichia coli and six serogroups of Shiga toxin-producing Escherichia coli, and 3 strains of Staphylococcus aureus and 10 serotypes of Salmonella; in \citet{hebling2022} an interesting review has been performed using different methods  (hyperspectral, NIR, Raman) in determine the adulteration in different milks, milks derivated (cheese, butter ) and other foot products and listing the unique works to the authors knowledge using visible together with Near Infra Red (NIR) images to determine adulteration in yogurt \citet{He2006StudyOL}.
In this work, we propose a combined analysis using the aforementioned devices and methodologies in order to determine the main properties of Cow's milk. 
Moreover, we also introduce visible imaging: it uses lenses and sensors optimized for the visible spectrum, typically across three bands (red, green, and blue), to create detailed images providing spatial information based on variations in light intensity and color: this enables the visualization of structures and surfaces under natural or controlled lighting conditions.

\section{Materials and Methods}
A total of 52 milk samples were taken from 26 cows housed at the Università Cattolica del Sacro Cuore dairy barn (CERZOO, San Bonico, Piacenza, Italy), in compliance with Italian regulations on animal experimentation and ethics, as described in \citet{Cattaneo2023}. 
Samples were taken at two time points: before dry-off (T0) and 35 days after calving (T12).

The healthy cows were randomly assigned to three experimental groups: (0) the SIG group (n = 9), dried off using only an internal teat sealant (Noroseal, Norbrook Laboratories Limited); (1) the control group (CTR; n = 9), dried off with a standard antibiotic treatment (Mamyzin, Boehringer Ing.Anim.H., Italy) and the internal teat sealant; and (2) the ASIG group (n = 8), which received the internal teat sealant supplemented with an oral administration of 200 mL/day of Aloe arborescens whole leaf homogenate from day -7 to day +7 relative to drying off.

These samples have been sub-sampled in two parts and stored in fridges at a constant temperature of twenty degrees Celsius below zero.

The first set of sub-samples was again sub-sampled twice, thawed, and sonified prior to spectrophotometric analysis for total polyphenol content and FRAP levels. In parallel, these samples were characterized via hyperspectral imaging. The second set, following thawing and sonication, was analyzed by GLC to determine fatty acid profiles, as detailed in Table \ref{tab:fattyacids}.
In all analyses, the samples were extracted using sterile graduated pipettes in predetermined volumes and placed into separate sterile transparent specimen containers.

\subsection{Gas-liquid chromatography}
The milk samples, subjected to lipid extraction, were dissolved in hexane as described by \citet{Cremonesi2024}. Methyl esters of medium- and long-chain fatty acids were prepared by an alkali-catalyzed transmethylation procedure with nonadecanoic acid methyl ester (Sigma Chemical Co., St. Louis, MO) as an internal standard. The composition of medium- and long-chain fatty acids was determined by gas chromatography using a ThermoQuest gas chromatograph (Milan, Italy) equipped with an FID and a high-polar fused silica capillary column (Chrompack CP-Sil 88 Varian, Middelburg, The Netherlands; 100 m × 0.25 mm i.d.; film thickness 0.20 mm). Helium was used as the carrier gas at a flow rate of 1 ml/min (\citet{CONTE20186497}). 

\begin{table}[ht!]
\tiny
\centering
\caption{Nomenclature and impact of fatty acids analysed}
\begin{tabular}{l p{5cm} p{5cm} p{5cm}}

\toprule
Nomenclature & Extended Name & Impact \\
\midrule
C4:0 & Butyric or Butanoic acid & Short-chain fatty acid (SCFA) vital for colon health; primary energy source for colonocytes \\
C6:0 & Caproic or Hexanoic acid& Short-chain fatty acid (SCFA); readily absorbed and used for energy \\
C8:0 & Caprylic or Octanoic acid & Medium-chain triglyceride (MCT); often used as a quick energy source and for antimicrobial properties \\
C10:1c9 & Capric or Decenoic acid with a cis double bond at position 9 & Less common fatty acid; contributes to the lipid profile of certain fats \\
C13:0ante & anteiso-Tridecanoic acid, branched at the $\omega$-2 position & A branched-chain fatty acid (BCFA); primarily used as a biomarker for specific bacterial activity (e.g., in ruminant fats) \\
C13:0 & Tridecanoic acid & Odd-chain fatty acid; considered a potential biomarker for dairy fat intake \\
C14:0iso & Iso-Myristic or iso-Tetradecanoic acid, branched at the $\omega$-1 position & Branched-chain fatty acid; reflects bacterial fermentation in the gut or diet \\
C14:0 & Myristic or Tetradecanoic acid) & Saturated fat; potentially raises LDL cholesterol (a cardiovascular risk factor) more significantly than C16:0 \\
C15:0iso & iso-Pentadecanoic acid (branched at the $\omega$-1 position) & Branched-chain fatty acid; reflects bacterial fermentation in the gut or diet \\
C14:1c9 & Myristoleic acid (or Tetradecenoic acid) with a cis double bond at position 9& Monounsaturated fatty acid (MUFA); contributes to cell membrane fluidity \\
C15:0 & Pentadecanoic acid & Odd-chain fatty acid; widely used as a biomarker for whole-milk dairy consumption \\
C16:0iso& iso-Palmitic acid (or iso-Hexadecanoic acid, branched at the $\omega$-1 position)& Branched-chain fatty acid; often indicates microbial or bacterial activity \\
C16:0 & Palmitic acid (or Hexadecanoic acid) & Most common saturated fat; high intake is associated with elevated cholesterol and cardiovascular risk \\
C17:0 & Margaric acid (or Heptadecanoic acid) & Odd-chain fatty acid; a useful biomarker for dairy fat intake. \\
C18:0 & Stearic acid (or Octadecanoic acid) & Saturated fat; considered neutral regarding cholesterol, it doesn't appear to raise LDL cholesterol \\
C18:1t10 & trans-Vaccenic acid (or Octadecenoic acid) with a trans double bond at position 10 & Ruminant trans fat; converts to CLA; generally considered less harmful than industrial trans fats \\
C18:1t16& Octadecenoic acid with a trans double bond at position 16& Less common trans fat; can be formed during industrial processing or from certain foods \\
C18:3n3 & Alpha-Linolenic acid (ALA) (an Omega-3 or n-3 Octadecatrienoic acid) & Essential Omega-3 fatty acid; precursor to EPA/DHA; vital for cardiovascular and brain health \\
C18:2c9t11 & Conjugated Linoleic Acid (CLA) with cis at 9 and trans at 11 & Naturally found in meat and dairy; studied for potential anti-cancer and weight management effects (mixed results) \\
C21:0 & Heneicosanoic acid & Long-chain saturated fat; low concentration makes its direct health impact minimal; often a component of waxes \\
C20:5n3 & Eicosapentaenoic acid (EPA) (an Omega-3 or n-3 Eicosapentaenoic acid) & Major Omega-3; powerful anti-inflammatory effects; essential for cardiovascular and immune function \\
\end{tabular}
\label{tab:fattyacids}
\end{table}


\subsection{Chemicals and Reagents}
\label{subsec1}

All standards and reagents were of analytical grade. Folin–Ciocalteu (FC) reagent, sodium carbonate (\ce{Na2}\ce{CO3}, sodium acetate (\ce{C2}\ce{H3}\ce{NaO2}, gallic acid, 2,4,6-tripyridyl-s-triazine (TPTZ), ferric chloride hexahydrate (\ce{FeCl3}·\ce{6H2}O), and ferrous sulfate heptahydrate (FeS\ce{O4}·7\ce{H2}O) were purchased from Sigma-Aldrich (St. Louis, MO, USA). Methanol was purchased from VWR (Radnor, PA, USA). Hydrochloric acid (HCl) was obtained from Merck (Darmstadt, Germany). Deionized water was used for all the preparations.

\subsection{Ferric Reducing Antioxidant Power (FRAP) Assay}
\label{subsec2}

The ferric reducing antioxidant power (FRAP) assay was performed following a modified version of the method described by \citet{BENZIE199915}. Fresh FRAP working solution was prepared by combining 10 volumes of 300 mM acetate buffer ($(p\mathrm{H}\, 3.6)$), 1 volume of 10 mM TPTZ dissolved in 40 mM HCl, and 1 volume of 20 mM Fe\ce{Cl3} 6H\ce {2O}. For the assay, \SI{735}{\micro\liter} of the reagent was mixed with \SI{35}{\micro\liter} of milk, and the reaction mixture was incubated at room temperature in the dark for 30 minutes. Absorbance was measured at 593 nm using a FLUOstar Omega Microplate Reader (BMG LABTECH, Ortenberg, Germany). Antioxidant capacity was expressed as mg of \ce{Fe^{2+}} equivalents per milliliters of milk (mg \ce{Fe^{2+}}/mL), using a calibration curve prepared with Fe\ce{SO4}·7H2\ce{O} (31.25–2000 µM).

\subsection{Determination of total polyphenol content}
\label{subsec3}

The total polyphenol content in cow milk samples was determined using the Folin-Ciocalteu colorimetric assay, following the procedure reported by \citet{singleton199914}, with minor modifications.
In brief, \SI{100}{\micro\liter} of milk was combined with \SI{500}{\micro\liter} of Folin-Ciocalteu reagent (diluted 1:10 in deionized water). After incubation in the dark at room temperature for 5 minutes, \SI{400}{\micro\liter} of \ce{Na2}\ce{CO3} solution (0.7 M) was added. The mixtures were further incubated under the same conditions for 2 hours. Finally, \SI{200}{\micro\liter} of each reaction mixture was transferred into a 96-well microplate, and absorbance was measured at 760 nm using a FLUOstar Omega Microplate Reader (BMG LABTECH, Germany).
Polyphenol content was expressed as milligrams of gallic acid equivalents (GAE) per milliliter (mg GAE/mL), calculated from a standard calibration curve prepared with gallic acid.

\subsection{Visible imaging}
\begin{figure}[ht!]
\centering
\includegraphics[width=0.6\textwidth]{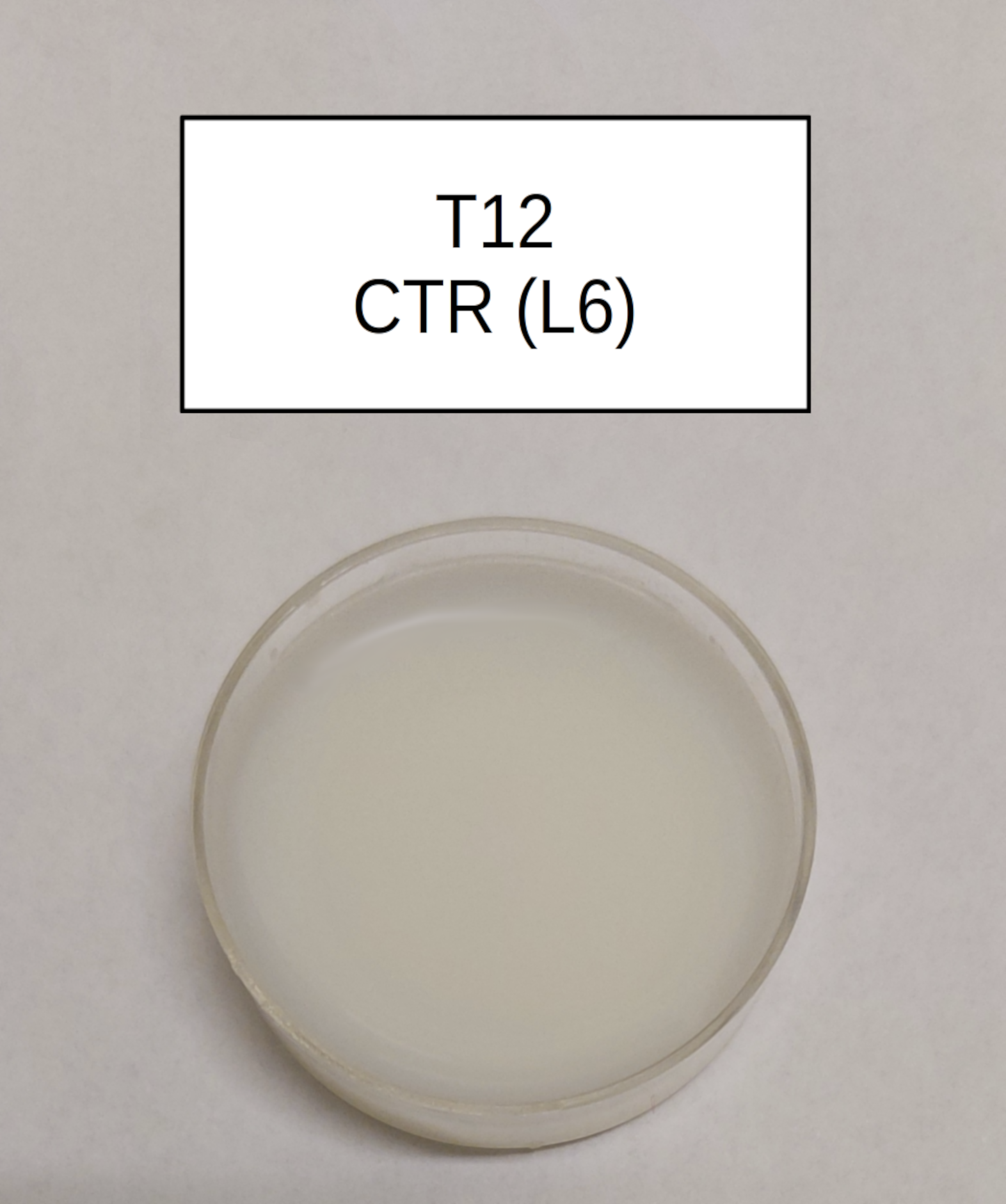}
\caption{Example of a picture of a milk sample taken with a mobile phone}\label{fig1}
\end{figure}

Images of the samples (Figure \ref{fig1}) were taken inside a laboratory with no windows at a constant temperature of 24 degrees Celsius and at a constant humidity, without direct light, under a lens hood at a variable brightness between 90 and 110 lumens using a mobile phone (Samsung A52S) camera set at ISO 125, speed 1/50, WV 3800 K, leaning on a support at a distance of 12 cm and activating the automatic shutter to prevent blurring. For each sample image a square of 512x512 pixels has been selected from the central part, then, for each square, first-order statistical description scales for color in images were calculated using Python numpy and Pillow libraries, specifically mean, standard deviation, texture contrast, texture energy, texture correlation, texture homogeneity and color histograms for the bins 293,301,365 and 366 where calculated from each RGB image.

\begin{figure}[ht!]
\centering
\includegraphics[width=0.8\textwidth]{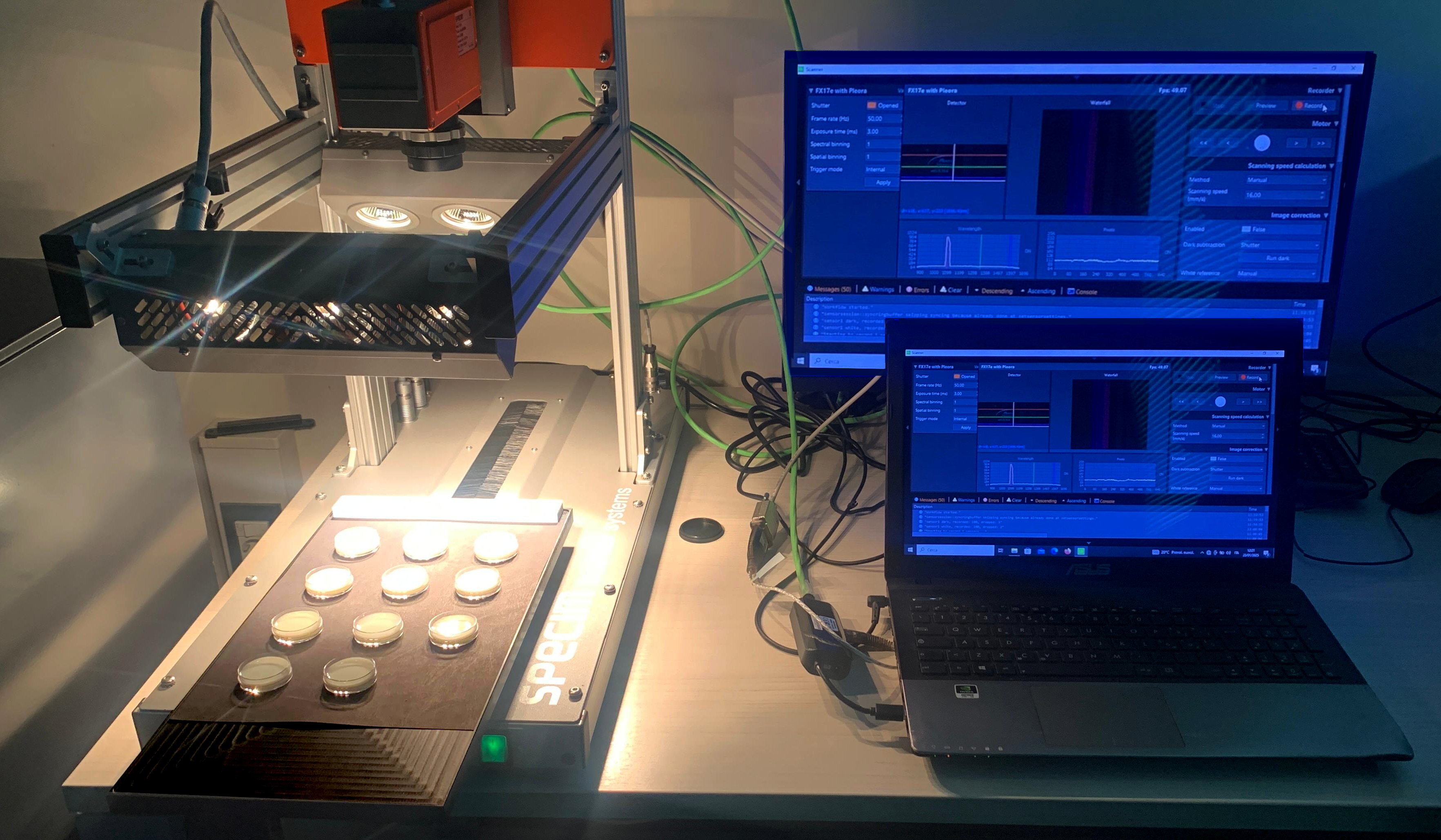}
\caption{Example of the preparation of the samples over the hyperspectral scanner tray}\label{fig2}
\end{figure}

Using different Machine Learning methods, that is, Spearman correlation and XGBoost, the features listed above have been compared with the primary milk features extracted using classical GLC devices.
Moreover, a subset of the same samples has been distributed over the tray (Figure \ref{fig2}) of a hyperspectral scanner (Specim FX17e).


Data from visible images, hyperspectral images (as for visible images, also in this case we selected squares of prefixed sizes from the central part of each sample), polyphenols, FRAP, and fatty acids were gathered and analysed using eleven different methods of correlation, regression, and classification:
\begin{enumerate}
\item the Pearson correlation coefficient was calculated to assess the linear relationship between the variables;
\item  the Kendall’s Tau coefficient was calculated to measure the strength and direction of the association between two ordinal or ranked variables;
\item the Principal Component Analysis (PCA) was performed to explain the cumulative variance in the dataset;
\item the classification with Support Vector Machine (SVM);
\item  the Grid Search for SVM;
\item  the classification with Random Forest;
\item  the classification with k-Nearest Neighbors (k-NN):
\item  the classification with a Multi Layer Perceptron (MLP) Neural Network;
\item  the regression with Partial Least Squares (PLS) specifically for hyperspectral data;
\item  the Long Short-Term Memory Regression for Hyperspectral data;
\item  the integration of LASSO, 1DCNN, and XGBoost for Hyperspectral data.
\end{enumerate}

While the first 10 methods are well-established analytical techniques, the 11th, \texttt{LASSO\_1DCNN\_XGB} uses an unsupervised clustering approach applied to hyperspectral data, grouping similar spectral profiles into homogeneous clusters without predefined labels. Initially, the Minimum Noise Fraction (MNF) transformation reduces data dimensionality while enhancing the signal-to-noise ratio, providing a cleaner, more informative dataset for subsequent analysis.
The Principal Component Analysis (PCA) is then applied to further reduce dimensionality by identifying the main directions of variance in the MNF-transformed data. This step helps simplify the data while retaining the most important spectral features.
Subsequently, the K-Means algorithm clusters the reduced data into distinct groups based on spectral similarity. The clustering quality is assessed using metrics such as the Silhouette Score to ensure well-defined, well-separated clusters.
To validate the clustering in terms of real-world significance, an Analysis of Variance (ANOVA) test is conducted between clusters on auxiliary variables associated with spectral profiles. The ANOVA p-value indicates whether the differences between groups are statistically significant, confirming that the grouping captures a meaningful structure in the data set.
This multi-step pipeline effectively segments complex hyperspectral datasets, ensuring that the resulting clusters are both mathematically robust and interpretable with respect to the properties analysed.

\section{Results}
The results are organized into three main sections: Exploratory Data Analysis, Visible Spectrum Analysis, and Hyperspectral Analysis. 

\subsection{Exploratory data analysis of milk bioactive components and antioxidant profiles}

First, an exploratory data analysis was conducted in order to 
characterise the milk samples,
highlighting the interaction between fatty acids, polyphenols, and FRAP.
Table \ref{tab:tab2} shows a summary of the mean values and standard deviations for each Cow group, while Table \ref{tab:tab3} shows a summary of the mean values and standard deviations for each experimental group at the two milking times (T0 and T12).

\begin{table}[ht!]
\centering
\begin{tabular}{l c c c}
\toprule
\multicolumn{1}{c}{} & \multicolumn{3}{c}{\textbf{GROUP}} \\
\cmidrule(lr){2-4}
Parameter & SIG = 0 & CTR = 1 & ASIG = 2 \\
\midrule
polyphenols & 1.32 $\pm$ 0.44 & 1.27 $\pm$ 0.44 & 1.39 $\pm$ 0.55 \\ 
frap           & 0.73 $\pm$ 0.47 & 0.63 $\pm$ 0.38 & 0.85 $\pm$ 0.60 \\ 
C4:0           & 1.41 $\pm$ 0.52 & 1.37 $\pm$ 0.25 & 1.63 $\pm$ 0.39 \\ 
C6:0 & 1.30 $\pm$ 0.24 & 1.30 $\pm$ 0.23 & 1.46 $\pm$ 0.24 \\ 
C8:0 & 1.55 $\pm$ 2.70 & 0.91 $\pm$ 0.27 & 1.68  $\pm$ 2.86 \\ 
C10:0 & 2.41 $\pm$ 0.33 & 2.39 $\pm$ 0.66 & 2.27 $\pm$ 0.51 \\
C10:1c9 & 0.27 $\pm$ 0.06 & 0.25 $\pm$ 0.08 & 0.27 $\pm$ 0.11 \\
C11:0 & 0.06 $\pm$ 0.02 & 0.06 $\pm$ 0.03 & 0.06 $\pm$ 0.02 \\
C12:0 & 3.16 $\pm$ 0.44 & 3.13 $\pm$ 0.78 & 2.95 $\pm$ 0.66 \\
C13:0iso & 0.04 $\pm$ 0.02 & 0.03 $\pm$ 0.01 & 0.05 $\pm$ 0.03 \\
C13:0ante & 0.09 $\pm$ 0.04 & 0.07 $\pm$ 0.04 & 0.07 $\pm$ 0.04 \\
C12:1c11 & 0.1 $\pm$ 0.07 & 0.09 $\pm$ 0.03 & 0.1 $\pm$ 0.03 \\
C13:0 & 0.1 $\pm$ 0.03 & 0.15 $\pm$ 0.08 & 0.09 $\pm$ 0.03 \\
C14:0iso & 0.14 $\pm$ 0.10 & 0.09 $\pm$ 0.05 & 0.14 $\pm$ 0.05 \\
C14:0 & 10.7 $\pm$ 0.94 & 10.72 $\pm$ 0.95 & 10.32 $\pm$ 1.29 \\
C15:0iso & 0.2 $\pm$ 0.07 & 0.21 $\pm$ 0.05 & 0.21 $\pm$ 0.05 \\
C15:0ante & 0.5 $\pm$ 0.15 & 0.46 $\pm$ 0.12 & 0.46 $\pm$ 0.10 \\
C14:1c9 & 1.04 $\pm$ 0.35 & 1.08 $\pm$ 0.26 & 1.08 $\pm$ 0.45 \\
C15:0 & 1.2 $\pm$ 0.13 & 1.3 $\pm$ 0.39 & 1.1 $\pm$ 0.18 \\
C16:0iso & 0.27 $\pm$ 0.09 & 0.25 $\pm$ 0.08 & 0.27 $\pm$ 0.08 \\
C17:0iso & 0.41 $\pm$ 0.10 & 0.43 $\pm$ 0.11 & 0.45 $\pm$ 0.08 \\
C16:1c7 & 0.09 $\pm$ 0.07 & 0.13 $\pm$ 0.09 & 0.1 $\pm$ 0.06 \\
C16:1c9 & 1.74 $\pm$ 0.32 & 1.94 $\pm$ 0.33 & 2.21 $\pm$ 0.50 \\
C17:0ante & 0.45 $\pm$ 0.06 & 0.46 $\pm$ 0.09 & 0.43 $\pm$ 0.06 \\
C17:0 & 0.59 $\pm$ 0.09 & 0.61 $\pm$ 0.14 & 0.59 $\pm$ 0.07 \\
C17:1c9 & 0.22 $\pm$ 0.07 & 0.29 $\pm$ 0.07 & 0.26 $\pm$ 0.07 \\
C18:0 & 9.75 $\pm$ 1.20 & 9.71 $\pm$ 1.58 & 9.16 $\pm$ 1.46 \\
C18:1t6-8 & 0.2 $\pm$ 0.09 & 0.21 $\pm$ 0.05 & 0.18 $\pm$ 0.04 \\ 
C18:1t9 & 0.13 $\pm$ 0.09 & 0.19 $\pm$ 0.07 & 0.15 $\pm$ 0.05 \\ 
C18:1t10 & 0.19 $\pm$ 0.1 & 0.39 $\pm$ 0.20 & 0.23 $\pm$ 0.12 \\ 
C18:1t11 & 0.57 $\pm$ 0.18 & 0.6 $\pm$ 0.20 & 0.56 $\pm$ 0.15 \\ 
C18:1t12 & 0.31 $\pm$ 0.12 & 0.34 $\pm$ 0.11 & 0.3 $\pm$ 0.10 \\  
C18:1c9 & 21.44 $\pm$ 3.47 & 22.46 $\pm$ 2.67 & 23.47 $\pm$ 3.48 \\ 
\end{tabular}
\end{table}


\begin{table}[ht!]
\centering
\begin{tabular}{l c c c}
\toprule
\multicolumn{1}{c}{} & \multicolumn{3}{c}{\textbf{GROUP}} \\
\cmidrule(lr){2-4}
Parameter & SIG = 0 & CTR = 1 & ASIG = 2 \\
\midrule
C18:1c11 & 0.71 $\pm$ 0.26 & 0.73 $\pm$ 0.22 & 0.65 $\pm$ 0.23 \\ 
C18:1c12 & 0.22 $\pm$ 0.05 & 0.23 $\pm$ 0.06 & 0.2 $\pm$ 0.04 \\
C18:1c13 & 0.06 $\pm$ 0.04 & 0.07 $\pm$ 0.05 & 0.08 $\pm$ 0.06 \\ 
C18:1t16 & 0.2 $\pm$ 0.05 & 0.21 $\pm$ 0.05 & 0.18 $\pm$ 0.03 \\ 
C18:2t9t12 & 0.05 $\pm$ 0.02 & 0.09 $\pm$ 0.03 & 0.07 $\pm$ 0.02 \\ 
C18:2t9c13 & 0.09 $\pm$ 0.04 & 0.13 $\pm$ 0.05 & 0.33 $\pm$ 0.49 \\
C18:2t8c12 & 0.1 $\pm$ 0.02 & 0.1 $\pm$ 0.05 & 0.07 $\pm$ 0.04 \\
C18:2t11c15 & 0.04 $\pm$ 0.02 & 0.04 $\pm$ 0.02 & 0.43 $\pm$ 0.52 \\
C18:2n-6 & 1.89 $\pm$ 0.28 & 2.05 $\pm$ 0.31 & 1.73 $\pm$ 0.6 \\
C20:0 & 0.1 $\pm$ 0.04 & 0.12 $\pm$ 0.04 & 0.13 $\pm$ 0.03 \\
C18:3n3 & 0.29 $\pm$ 0.07 & 0.37 $\pm$ 0.09 & 0.36 $\pm$ 0.08 \\
C18:2c9t11 & 0.31 $\pm$ 0.09 & 0.3 $\pm$ 0.13 & 0.25 $\pm$ 0.13 \\
C21:0 & 0.03 $\pm$ 0.01 & 0.04 $\pm$ 0.02 & 0.03 $\pm$ 0.02 \\
C20:2n6 & 0.03 $\pm$ 0.01 & 0.05 $\pm$ 0.05 & 0.11 $\pm$ 0.11 \\
C22:0 & 0.03 $\pm$ 0.01 & 0.06 $\pm$ 0.04 & 0.06 $\pm$ 0.01 \\
C20:3n6 & 0.13 $\pm$ 0.07 & 0.11 $\pm$ 0.03 & 0.11 $\pm$ 0.06 \\
C20:3n3 & 0.1 $\pm$ 0.07 & 0.08 $\pm$ 0.05 & 0.07 $\pm$ 0.05 \\
C20:4n6 & 0.17 $\pm$ 0.02 & 0.13 $\pm$ 0.04 & 0.18 $\pm$ 0.06 \\
C20:5n3 & 0.04 $\pm$ 0.03 & 0.04 $\pm$ 0.02 & 0.04 $\pm$ 0.01 \\
C24:0 & 0.04 $\pm$ 0.01 & 0.04 $\pm$ 0.01 & 0.07 $\pm$ 0.04 \\ 
C22:5n3 & 0.03 $\pm$ 0.01 & 0.09 $\pm$ 0.06 & 0.04 $\pm$ 0.01 \\
FA\_SAT & 72.55 $\pm$ 2.56 & 65.23 $\pm$ 2.57 & 72.5  $\pm$ 2.66 \\ 
FA\_MONO & 24.37 $\pm$ 7.21 & 27.28 $\pm$ 2.67 & 26.51  $\pm$ 7.96 \\ 
FA\_POLY & 2.53 $\pm$ 0.72 & 2.9 $\pm$ 0.42 & 2.47  $\pm$ 0.91 \\ 
OMEGA6 & 2.09 $\pm$ 0.61 & 2.33 $\pm$ 0.34 & 1.99  $\pm$ 0.77 \\ 
OMEGA3 & 0.44 $\pm$ 0.14 & 0.57 $\pm$ 0.14 & 0.48  $\pm$ 0.17 \\ 
OMEGA6\_3 & 4.9 $\pm$ 1.27 & 4.26 $\pm$ 1.05 & 4.19  $\pm$ 1.15 \\ 
\end{tabular}

\caption{Descriptive statistics of polyphenols, FRAP, and fatty acids across experimental groups and milking times}\label{tab:tab2}
\end{table}

\begin{figure}[ht!]
\centering
\begin{minipage}{0.48\textwidth}
\centering
\includegraphics[width=\textwidth]{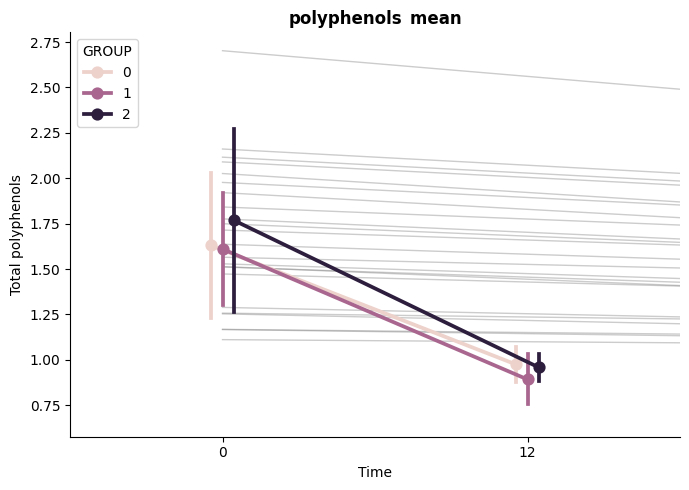}
\label{figfrapt0t12}
\end{minipage}\hfill
\begin{minipage}{0.48\textwidth}
\centering
\includegraphics[width=\textwidth]{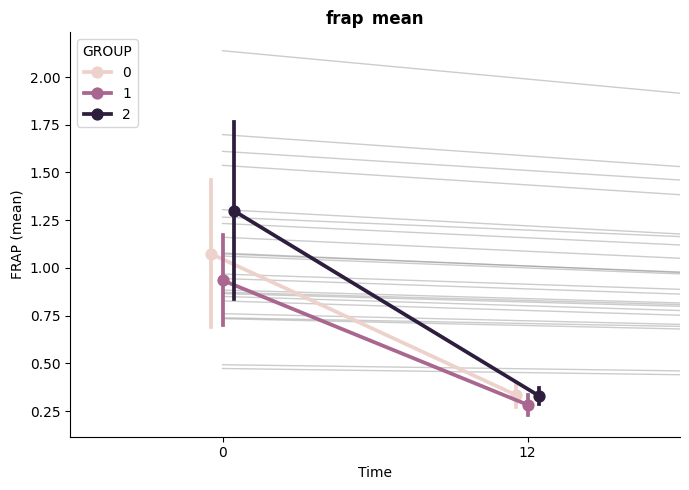}
\label{figpolyt0t12}
\end{minipage}
\vspace{-24pt}
\caption{Visual distribution of polyphenol and FRAP values at T0 and T12, highlighting the shift between the two milking times}
\label{fig:frappolyt012}
\end{figure}

\begin{figure}[ht!]
\centering
\begin{minipage}{0.48\textwidth}
\centering
\includegraphics[width=\textwidth]{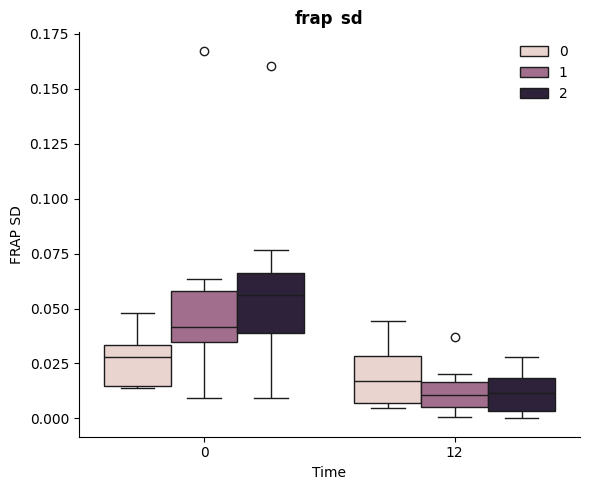}
\label{figfrapsdt0t12}
\end{minipage}\hfill
\begin{minipage}{0.48\textwidth}
\centering
\includegraphics[width=\textwidth]{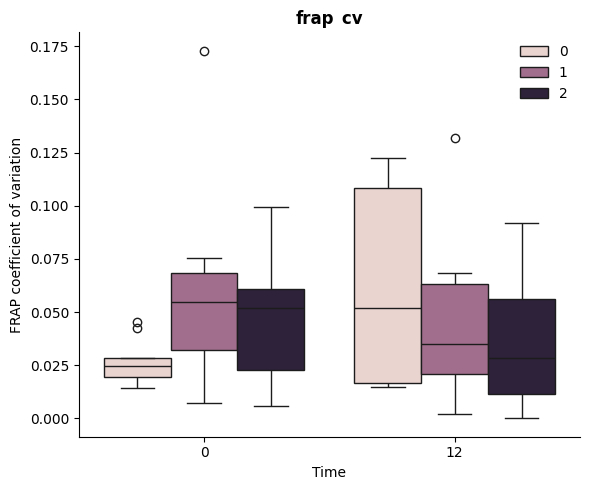}
\label{figfrapcvt0t12}
\end{minipage}
\vspace{-24pt}
\caption{Boxplots illustrating the standard deviations and the coefficient of variation of FRAP. The data show that the treatments significantly influence the variability of the antioxidant response rather than the absolute magnitude of the parameters}
\label{fig:frapsdcvt0t12}
\end{figure}

The exploratory data analysis of antioxidant capacity reveals a sharp decline between the two milking sessions (T0 vs T12) across all experimental groups. Distinct distribution patterns are also evident among the cow groups (Figure \ref{fig:frappolyt012}). Moreover, the analysis of variance and data dispersion, expressed as the standard deviation (sd) and the coefficient of variation (cv), confirm that the experimental treatments significantly modulate the variance of FRAP levels (Figure \ref{fig:frapsdcvt0t12}).

\subsection{Visible analysis}
\begin{figure}[ht!]
\centering
\begin{minipage}{0.48\textwidth}
\centering
\includegraphics[width=\textwidth]{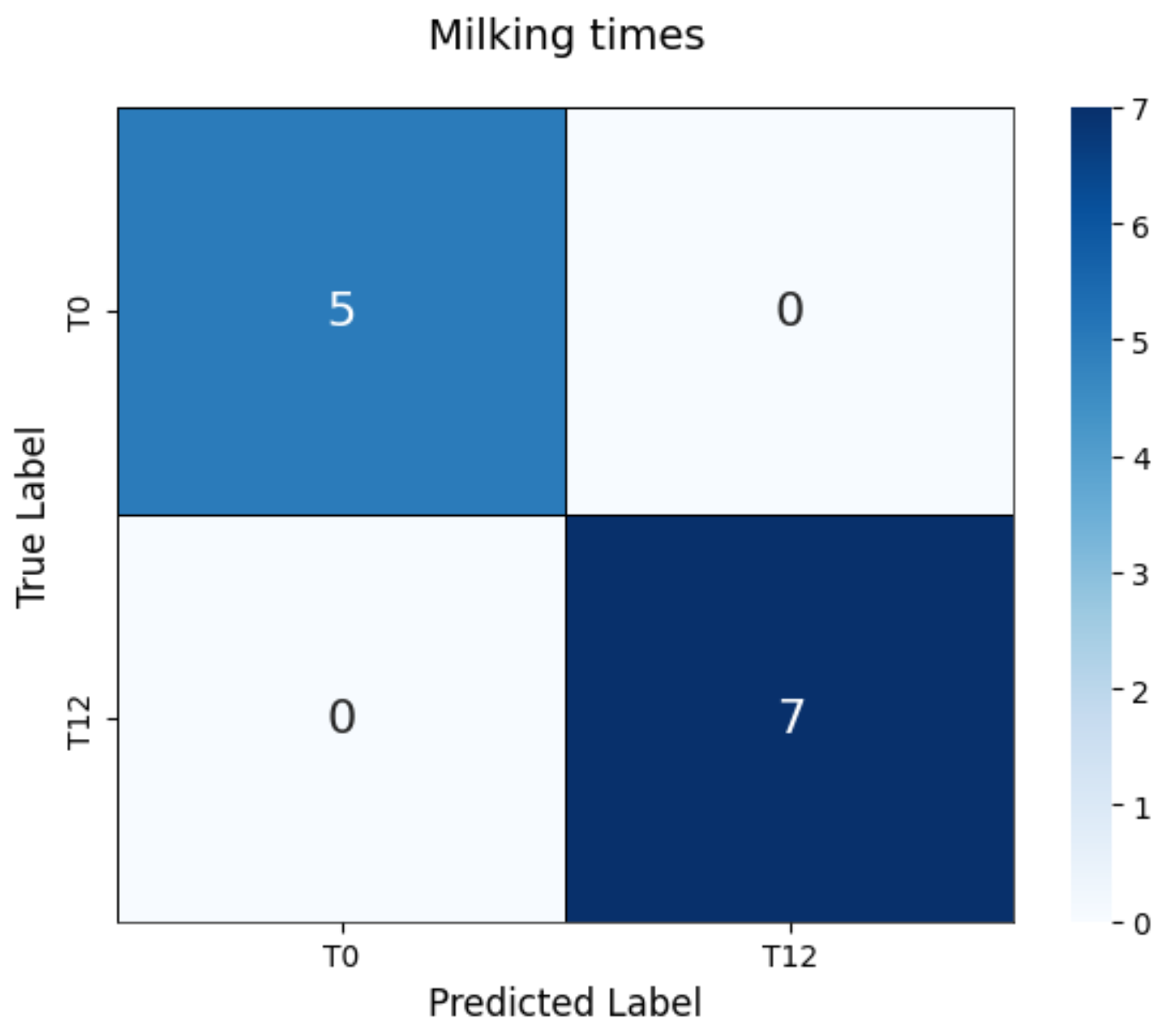}
\label{imgT0T12}
\end{minipage}\hfill
\begin{minipage}{0.48\textwidth}
\centering
\includegraphics[width=\textwidth]{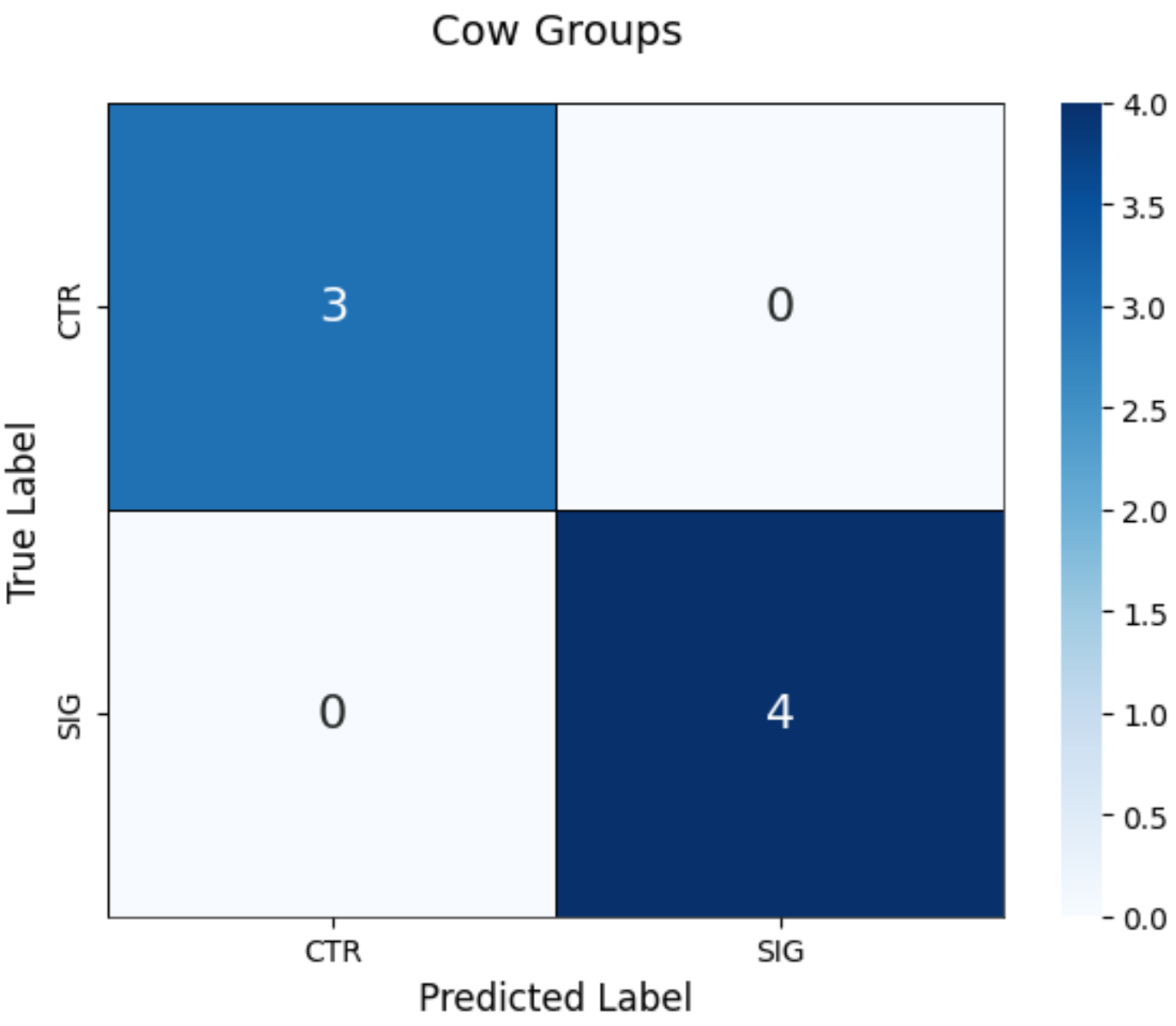}
\label{imgCTRSIG}
\end{minipage}
\caption{Confusion matrix of the test dataset for image classification of milking time (T0, T12) and of CTR, SIG. Identical results achieved with a decision tree classifier, an SVM, and a three-layer feed-forward perceptron}
\label{fig:imgt0t12CTRSIG}
\end{figure}

\begin{figure}[ht!]
\centering
\begin{minipage}{0.48\textwidth}
\centering
\includegraphics[width=\textwidth]{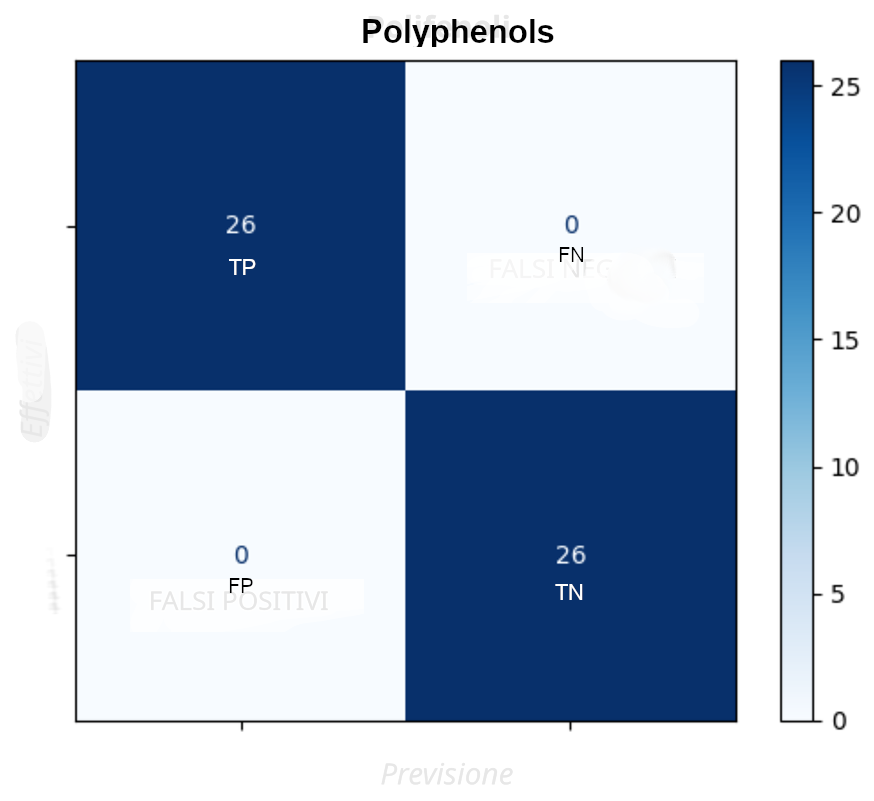} 
\label{cmimgpoly}
\end{minipage}\hfill
\begin{minipage}{0.48\textwidth}
\centering
\includegraphics[width=\textwidth]{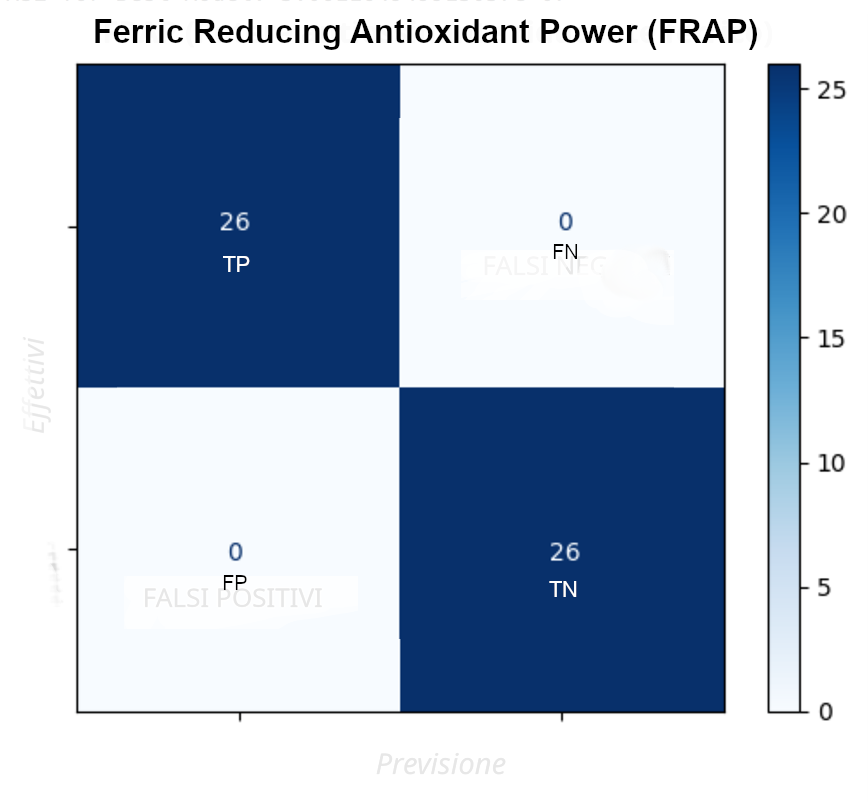} 
\label{cmimfrap}
\end{minipage}
\caption{Confusion matrix for image classification of Polyphenols and FRAP obtained using XGBBoost method}
\label{fig:cmimagepolyfrap}
\end{figure}

The 52 images acquired through the smartphone camera enabled us to:
\begin{itemize}
    \item distinguishes between samples obtained at T0 and T12 with 100\% accuracy (both for training and for testing); the dataset was randomized and split into training and testing sets (80\% and 20\%, respectively), then analysed using a DecisionTreeClassifier (scikit-learn), a Support Vector Machine (SVM), and a three-layer Feed-Forward Perceptron with two hidden layers (64 and 32 neurons) and a single sigmoid output) running 3 epochs, all of which yielded identical results (Figure \ref{fig:imgt0t12CTRSIG}:\ref{imgT0T12};
    \item distinguish CTR from SIG samples, using the same methods, achieving an accuracy of 100\% (Figure \ref{fig:imgt0t12CTRSIG}:\ref{imgT0T12};
    \item in addition, a clear significance is the correlation of visible images and polyphenols, and also from visible images and the mean FRAP, with an accuracy of 100\% using the XGBoost method (Figure \ref{fig:cmimagepolyfrap}).
\end{itemize}

\begin{table}[ht!]
\tiny
\centering
\begin{tabular}{l c c c c c c}
\toprule
 & \multicolumn{3}{c}{TIME 0} & \multicolumn{3}{c}{TIME 12} \\
\cmidrule(lr){2-4} \cmidrule(lr){5-7}
Parameter & SIG & CTR & ASIG & SIG & CTR & ASIG \\
\midrule
polyphenols & $1.63 \pm 0.40$ & $1.61 \pm 0.31$ & $1.77 \pm 0.50$ & $0.97 \pm 010$ & $0.89 \pm 0.14$ & $0.96 \pm 0.08$ \\
frap & $1.07 \pm 0.39$ & $0.94 \pm 0.24$ & $1.30 \pm 0.46$ & $0.33 \pm 0.06$ & $0.28 \pm 0.05$ & $0.33 \pm 0.04$ \\
C4:0 & 1.44 $\pm$ 0.62 & 1.25 $\pm$ 0.26 & 1.56 $\pm$ 0.46 & 1.37 $\pm$ 0.41 & 1.49 $\pm$ 0.16 & 1.72 $\pm$ 0.27 \\
C6:0 & 1.34 $\pm$ 0.27 & 1.35 $\pm$ 0.18 & 1.51 $\pm$ 0.21 & 19.47 $\pm$ 1.53 & 1.24 $\pm$ 0.31 & 26.57 $\pm$ 6.78 \\
C8:0 & 0.89 $\pm$ 0.28 & 0.9 $\pm$ 0.3 & 1.01 $\pm$ 0.23 & 2.3 $\pm$ 3.92 & 0.92 $\pm$ 0.24 & 2.45 $\pm$ 4.21 \\
C10:0 & 2.42 $\pm$ 0.21 & 2.42 $\pm$ 0.64 & 2.43 $\pm$ 0.58 & 2.39 $\pm$ 0.45 & 2.35 $\pm$ 0.73 & 2.06 $\pm$ 0.32 \\
C10:1c9 & 0.31 $\pm$ 0.03 & 0.28 $\pm$ 0.07 & 0.32 $\pm$ 0.11 & 0.22 $\pm$ 0.03 & 0.22 $\pm$ 0.07 & 0.19 $\pm$ 0.04 \\
C11:0 & 0.06 $\pm$ 0.02 & 0.05 $\pm$ 0.03 & 0.07 $\pm$ 0.02 & 0.06 $\pm$ 0.02 & 0.06 $\pm$ 0.02 & 0.04 $\pm$ 0.0 \\
C12:0 & 3.26 $\pm$ 0.34 & 3.22 $\pm$ 0.8 & 3.31 $\pm$ 0.57 & 3.03 $\pm$ 0.54 & 3.03 $\pm$ 0.79 & 2.46 $\pm$ 0.39 \\
C13:0iso & 0.03 $\pm$ 0.01 & 0.04 $\pm$ 0.01 & 0.06 $\pm$ 0.04 & 0.05 $\pm$ 0.03 & 0.03 $\pm$ 0.01 & 0.04 $\pm$ 0.01 \\
C13:0ante & 0.11 $\pm$ 0.02 & 0.07 $\pm$ 0.04 & 0.11 $\pm$ 0.03 & 0.06 $\pm$ 0.03 & 0.08 $\pm$ 0.03 & 0.03 $\pm$ 0.0 \\
C12:1c11 & 0.09 $\pm$ 0.03 & 0.1 $\pm$ 0.04 & 0.11 $\pm$ 0.03 & 0.1 $\pm$ 0.11 & 0.08 $\pm$ 0.03 & 0.08 $\pm$ 0.03 \\
C13:0 & 0.11 $\pm$ 0.02 & 0.12 $\pm$ 0.05 & 0.11 $\pm$ 0.03 & 0.1 $\pm$ 0.03 & 0.18 $\pm$ 0.1 & 0.06 $\pm$ 0.01 \\
C14:0iso & 0.12 $\pm$ 0.05 & 0.12 $\pm$ 0.04 & 0.14 $\pm$ 0.02 & 0.16 $\pm$ 0.14 & 0.06 $\pm$ 0.03 & 0.14 $\pm$ 0.08 \\
C14:0 & 10.86 $\pm$ 0.61 & 10.91 $\pm$ 0.85 & 11.04 $\pm$ 1.02 & 10.5 $\pm$ 1.07 & 10.49 $\pm$ 1.28 & 9.36 $\pm$ 0.96 \\
C15:0iso & 0.22 $\pm$ 0.06 & 0.24 $\pm$ 0.05 & 0.25 $\pm$ 0.03 & 0.17 $\pm$ 0.07 & 0.17 $\pm$ 0.02 & 0.17 $\pm$ 0.04 \\ 
C15:0ante & 0.52 $\pm$ 0.06 & 0.55 $\pm$ 0.07 & 0.53 $\pm$ 0.04 & 0.49 $\pm$ 0.22 & 0.36 $\pm$ 0.08 & 0.37 $\pm$ 0.06 \\ 
C14:1c9 & 1.25 $\pm$ 0.23 & 1.21 $\pm$ 0.26 & 1.36 $\pm$ 0.33 & 0.77 $\pm$ 0.28 & 0.93 $\pm$ 0.19 & 0.72 $\pm$ 0.3 \\ 
C15:0 & 1.21 $\pm$ 0.13 & 1.22 $\pm$ 0.22 & 1.2 $\pm$ 0.13 & 1.19 $\pm$ 0.14 & 1.38 $\pm$ 0.53 & 0.96 $\pm$ 0.14 \\
C16:0iso & 0.3 $\pm$ 0.11 & 0.31 $\pm$ 0.06 & 0.32 $\pm$ 0.04 & 0.24 $\pm$ 0.06 & 0.19 $\pm$ 0.05 & 0.2 $\pm$ 0.07\\
C16:0 & 35.61 $\pm$ 4.16 & 33.18 $\pm$ 3.4 & 34.51 $\pm$ 1.34 & 35.09 $\pm$ 2.05 & 33.5 $\pm$ 2.25 & 31.61 $\pm$ 0.55 \\
C17:0iso & 0.41 $\pm$ 0.11 & 0.41 $\pm$ 0.1 & 0.39 $\pm$ 0.04 & 0.40 $\pm$ 0.09 & 0.45 $\pm$ 0.13 & 0.52 $\pm$ 0.04 \\
C16:1c7 & 0.1 $\pm$ 0.06 & 0.17 $\pm$ 0.09 & 0.14 $\pm$ 0.06 & 0.08 $\pm$ 0.08 & 0.09 $\pm$ 0.07 & 0.04 $\pm$ 0.0 \\
C16:1c9 & 1.72 $\pm$ 0.31 & 1.95 $\pm$ 0.34 & 2.24 $\pm$ 0.62 & 1.77 $\pm$ 0.36 & 1.93 $\pm$ 0.35 & 2.16 $\pm$ 0.33 \\
C17:0ante & 0.46 $\pm$ 0.06 & 0.49 $\pm$ 0.08 & 0.46 $\pm$ 0.07 & 0.43 $\pm$ 0.06 & 0.41 $\pm$ 0.08 & 0.4 $\pm$ 0.05 \\
C17:0 & 0.55 $\pm$ 0.07 & 0.62 $\pm$ 0.1 & 0.53 $\pm$ 0.04 & 0.65 $\pm$ 0.08 & 0.61 $\pm$ 0.18 & 0.66 $\pm$ 0.03 \\
C17:1c9 & 0.21 $\pm$ 0.05 & 0.28 $\pm$ 0.07 & 0.23 $\pm$ 0.05 & 0.24 $\pm$ 0.08 & 0.3 $\pm$ 0.06 & 0.3 $\pm$ 0.09 \\
C18:0 & 9.49 $\pm$ 1.51 & 9.6 $\pm$ 1.62 &8.52 $\pm$ 1.6 & 10.08 $\pm$ 0.56 & 9.82 $\pm$ 1.63 & 10.03 $\pm$ 0.63 \\
C18:1t6-8 & 0.22 $\pm$ 0.1 & 0.24 $\pm$ 0.04 & 0.17 $\pm$ 0.04 & 0.18 $\pm$ 0.08 & 0.17 $\pm$ 0.04 & 0.18 $\pm$ 0.04 \\
C18:1t9 & 0.12 $\pm$ 0.06 & 0.2 $\pm$ 0.03 & 0.15 $\pm$ 0.05 & 0.15 $\pm$ 0.12 & 0.17 $\pm$ 0.1 & 0.15 $\pm$ 0.05 \\
C18:1t10 & 0.19 $\pm$ 0.11 & 0.4 $\pm$ 0.2 & 0.26 $\pm$ 0.14 & 0.19 $\pm$ 0.11 & 0.38 $\pm$ 0.22 & 0.18 $\pm$ 0.09 \\
C18:1t11 & 0.56 $\pm$ 0.16 & 0.64 $\pm$ 0.2 & 0.5 $\pm$ 0.1 & 0.59 $\pm$ 0.21 & 0.55 $\pm$ 0.22 & 0.64 $\pm$ 0.17 \\
C18:1t12 & 0.31 $\pm$ 0.11 & 0.4 $\pm$ 0.08 & 0.34 $\pm$ 0.05 & 0.31 $\pm$ 0.14 & 0.26 $\pm$ 0.08 & 0.25 $\pm$ 0.12 \\
C18:1c9 & 20.85 $\pm$ 2.84 & 21.77 $\pm$ 2.36 & 21.24 $\pm$ 2.31 & 22.2 $\pm$ 4.26 & 23.23 $\pm$ 2.94 & 26.44 $\pm$ 2.35 \\
C18:1c11 & 0.52 $\pm$ 0.14 & 0.64 $\pm$ 0.19 & 0.47 $\pm$ 0.07 & 0.96 $\pm$ 0.14 & 0.83 $\pm$ 0.21 & 0.89 $\pm$ 0.07 \\
C18:1c12 & 0.22 $\pm$ 0.05 & 0.27 $\pm$ 0.04 & 0.22 $\pm$ 0.03 & 0.20 $\pm$ 0.04 & 0.19 $\pm$ 0.06 & 0.17 $\pm$ 0.03 \\
C18:1c13 & 0.06 $\pm$ 0.03 & 0.07 $\pm$ 0.05 & 0.06 $\pm$ 0.08 & 0.08 $\pm$ 0.05 & 0.08 $\pm$ 0.05 & 0.10 $\pm$ 0.03 \\
C18:1t16 & 0.20 $\pm$ 0.05 & 0.24 $\pm$ 0.03 & 0.18 $\pm$ 0.04 & 0.18 $\pm$ 0.05 & 0.17 $\pm$ 0.02 & 0.18 $\pm$ 0.03 \\
C18:2t9t12 & 0.06 $\pm$ 0.03 & 0.09 $\pm$ 0.03 & 0.08 $\pm$ 0.03 & 0.03 $\pm$ 0.01 & 0.09 $\pm$ 0.04 & 0.07 $\pm$ 0.02 \\
C18:2t9c13 & 0.1 $\pm$ 0.05 & 0.15 $\pm$ 0.04 & 0.12 $\pm$ 0.03 & 0.09 $\pm$ 0.02 & 0.09 $\pm$ 0.03 & 0.6 $\pm$ 0.69 \\
C18:2t8c12 & 0.1 $\pm$ 0.03 & 0.14 $\pm$ 0.03 & 0.09 $\pm$ 0.05 & 0.09 $\pm$ 0.01 & 0.06 $\pm$ 0.03 & 0.05 $\pm$ 0.02 \\
C18:2t11c15 & 0.04 $\pm$ 0.02 & 0.05 $\pm$ 0.02 & 0.04 $\pm$ 0.01 & 0.03 $\pm$ 0.01 & 0.04 $\pm$ 0.02 & 0.97 $\pm$ 0.33 \\
C18:2n-6 & 1.98 $\pm$ 0.23 & 2.05 $\pm$ 0.29 & 2.09 $\pm$ 0.18 & 1.78 $\pm$ 0.32 & 2.05 $\pm$ 0.36 & 1.25 $\pm$ 0.64 \\
C20:0 & 0.11 $\pm$ 0.04 & 0.14 $\pm$ 0.03 & 0.12 $\pm$ 0.03 & 0.08 $\pm$ 0.02 & 0.10 $\pm$ 0.03 & 0.14 $\pm$ 0.0 \\
C18:3n3 & 0.29 $\pm$ 0.09 & 0.39 $\pm$ 0.09 & 0.42 $\pm$ 0.04 & 0.29 $\pm$ 0.03 & 0.34 $\pm$ 0.08 & 0.28 $\pm$ 0.04 \\ 
C18:2c9t11 & 0.36 $\pm$ 0.09 & 0.37 $\pm$ 0.11 & 0.33 $\pm$ 0.11 & 0.25 $\pm$ 0.06 & 0.22 $\pm$ 0.1 & 0.15 $\pm$ 0.08 \\ 
C21:0 & 0.04 $\pm$ 0.01 & 0.05 $\pm$ 0.02 & 0.05 $\pm$ 0.0 & 0.02 $\pm$ 0.0 & 0.04 $\pm$ 0.02 & 0.01 $\pm$ 0.0 \\ 
C20:2n6 & 0.04 $\pm$ 0.0 & 0.02 $\pm$ 0.01 & 0.03 $\pm$ 0.03 & 0.02 $\pm$ 0.0 & 0.08 $\pm$ 0.07 & 0.20 $\pm$ 0.11 \\ 
C22:0 & 0.03 $\pm$ 0.01 & 0.07 $\pm$ 0.04 & 0.06 $\pm$ 0.00 & 0.02 $\pm$ 0.01 & 0.05 $\pm$ 0.04 & 0.07 $\pm$ 0.01 \\ 
C20:3n6 & 0.13 $\pm$ 0.07 & 0.12 $\pm$ 0.03 & 0.14 $\pm$ 0.05 & 0.12 $\pm$ 0.08 & 0.09 $\pm$ 0.02 & 0.07 $\pm$ 0.04 \\ 
C20:3n3 & 0.13 $\pm$ 0.07 & 0.10 $\pm$ 0.06 & 0.07 $\pm$ 0.07 & 0.07 $\pm$ 0.04 & 0.06 $\pm$ 0.03 & 0.08 $\pm$ 0.01 \\ 
C20:4n6 & 0.19 $\pm$ 0.02 & 0.13 $\pm$ 0.03 & 0.17 $\pm$ 0.03 & 0.15 $\pm$ 0.02 & 0.12 $\pm$ 0.04 & 0.20 $\pm$ 0.09 \\ 
C20:5n3 & 0.03 $\pm$ 0.00 & 0.03 $\pm$ 0.00 & 0.03 $\pm$ 0.0 & 0.05 $\pm$ 0.04 & 0.05 $\pm$ 0.03 & 0.05 $\pm$ 0.01 \\ 
C24:0 & 0.03 $\pm$ 0.0 & 0.04 $\pm$ 0.01 & 0.06 $\pm$ 0.03 & 0.04 $\pm$ 0.01 & 0.04 $\pm$ 0.02 & 0.09 $\pm$ 0.03 \\ 
C22:5n3 & 0.03 $\pm$ 0.01 & 0.10 $\pm$ 0.08 & 0.05 $\pm$ 0.01 & 0.04 $\pm$ 0.01 & 0.07 $\pm$ 0.04 & 0.03 $\pm$ 0.01 \\ 
FA\_SAT & 67.44 $\pm$ 3.34 & 65.16 $\pm$ 2.43 & 66.1 $\pm$ 2.51 & 78.3 $\pm$ 3.84 & 65.31 $\pm$ 2.88 & 79.81 $\pm$ 4.63 \\ 
FA\_MONO & 25.33 $\pm$ 2.85 & 26.74 $\pm$ 2.43 & 26.39 $\pm$ 2.02 & 23.28 $\pm$ 1.34 & 27.88 $\pm$ 2.97 & 26.65 $\pm$ 1.96 \\ 
FA\_POLY & 2.81 $\pm$ 0.27 & 2.94 $\pm$ 0.39 & 3.00 $\pm$ 0.3 & 2.21 $\pm$ 0.94 & 2.86 $\pm$ 0.48 & 1.86 $\pm$ 1.01 \\ 
OMEGA6 & 2.33 $\pm$ 0.26 & 2.32 $\pm$ 0.31 & 2.43 $\pm$ 0.21 & 1.81 $\pm$ 0.78 & 2.34 $\pm$ 0.38 & 1.48 $\pm$ 0.89 \\ 
OMEGA3 & 0.48 $\pm$ 0.10 & 0.62 $\pm$ 0.13 & 0.57 $\pm$ 0.10 & 0.40 $\pm$ 0.17 & 0.52 $\pm$ 0.13 & 0.38 $\pm$ 0.17 \\ 
OMEGA6\_3 & 5.17 $\pm$ 1.51 & 3.88 $\pm$ 0.86 & 4.34 $\pm$ 0.51 & 4.55 $\pm$ 0.84 & 4.68 $\pm$ 1.14 & 4.0 $\pm$ 1.73 \\

\bottomrule
\end{tabular}
\caption{Parameters per GROUP per TIME}\label{tab2}
\end{table}

\begin{figure}[ht!]
\centering
\includegraphics[width=0.8\textwidth]{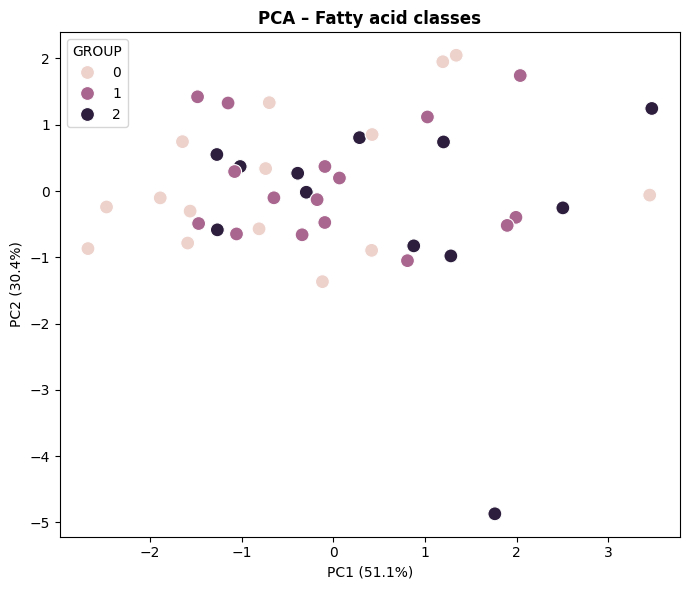}
\caption{Distribution of fatty acid classes across Groups 0, 1, and 2. High inter-group overlap indicates no significant clustering or discrimination based on the analyzed lipid profiles. PC1 and PC2 explain 51.1\% and 30.4\% of the total variance, respectively}\label{fig:pcafattyacidclasses}
\end{figure}

A Principal Component Analysis (PCA) was performed to evaluate the distribution of fatty acid classes across the study groups. As illustrated in Figure \ref{fig:pcafattyacidclasses}, the individual samples from the cow groups (SIG, CTR, and ASIG) show a significant overlap. The lack of distinct clustering suggests that the lipidic profiles are relatively homogenous across these categories, with no significant discrimination provided by the first two principal components.

\begin{figure}[ht!]
\centering
\includegraphics[width=0.8\textwidth]{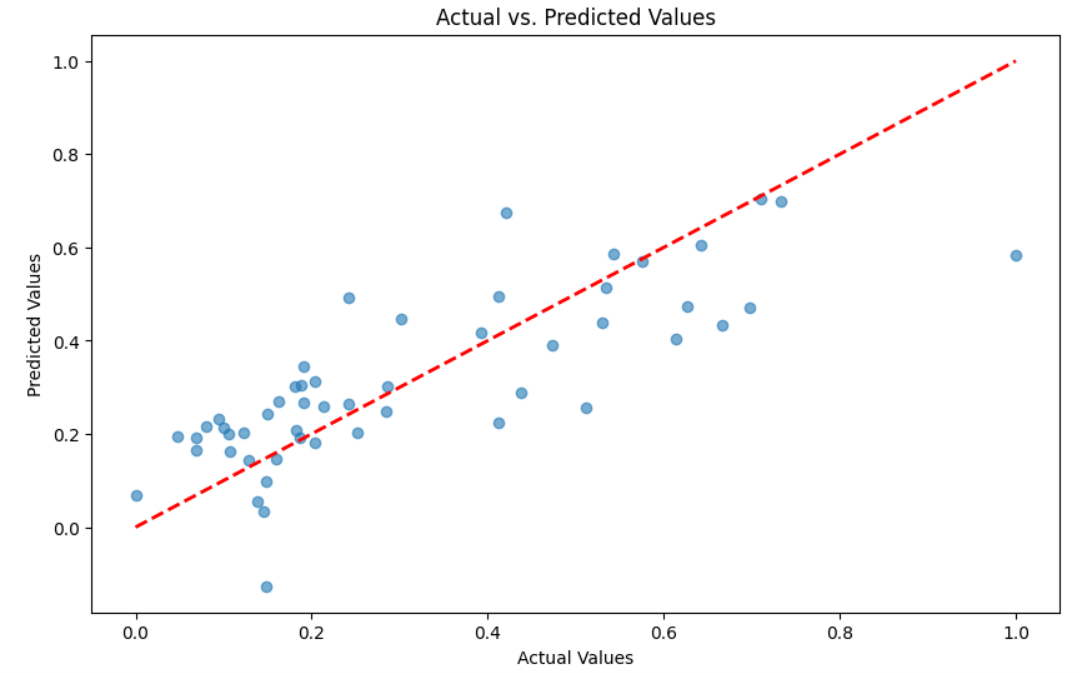}
\caption{Multivariate linear Regression plot of actual versus predicted values (Ordinary Least Squares) between the polyphenols mean and the values Texture contrast, Texture homogeneity, Texture energy, and Mean color channel 2}\label{fig:multivar1}
\end{figure}

\begin{figure}[ht!]
\centering
\includegraphics[width=0.8\textwidth]{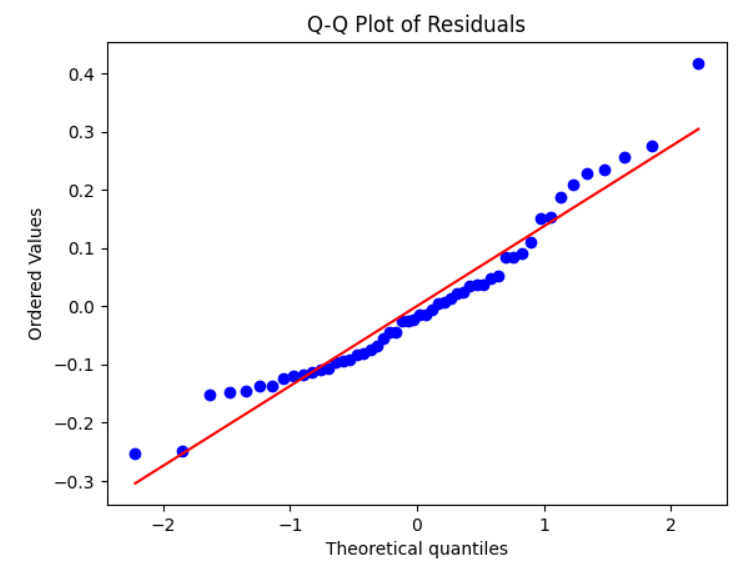}
\caption{Q-Q plot of the residuals of the linear Regression plot of actual versus predicted values (Ordinary Least Squares) between the polyphenols mean and the values Texture contrast, Texture homogeneity, Texture energy, and Mean color channel 2}\label{fig:qq1}
\end{figure}

In Output \ref{verb:olsreg1} a multivariate linear regression (Ordinary Least Squares) shows a correlation between the mean value of polyphenols mean and the values Texture contrast, Texture homogeneity, Texture energy and Mean color channel 2, supported by the Actual vs Predicted values image in Figure \ref{fig:multivar1} and by the related Q-Q plot of the residuals in Figure \ref{fig:qq1} showing the alignment along the diagonal line, indicating that the distribution of the residuals closely adheres to the theoretical distribution and telling that the model is appropriate and that the errors are random and not systematic.

\vspace{2\baselineskip}

{\scriptsize
\begin{labelverb}\label{verb:olsreg1}
\begin{verbatim}
OLS Regression Results                            
=====================================================================================
Dep. Variable:       polyphenols_mean          R-squared:                       0.635
Model:                            OLS          Adj. R-squared:                  0.604
Method:                 Least Squares          F-statistic:                     20.41
Date:                Sun, 09 Mar 2025          Prob (F-statistic):           8.41e-10
Time:                        16:13:18          Log-Likelihood:                 29.967
No. Observations:                  52          AIC:                            -49.93
Df Residuals:                      47          BIC:                            -40.18
Df Model:                           4                                         
Covariance Type:            nonrobust                                         
=====================================================================================
                        coef    std err          t      P>|t|      [0.025      0.975]
-------------------------------------------------------------------------------------
const                 0.3127      0.020     15.764      0.000       0.273       0.353
Texture contrast      0.7909      0.236      3.349      0.002       0.316       1.266
Texture homogeneity  -0.7127      0.215     -3.318      0.002      -1.145      -0.281
Texture energy        0.4678      0.122      3.842      0.000       0.223       0.713
Mean color channel 2  0.3009      0.100      3.012      0.004       0.100       0.502
=====================================================================================
Omnibus:                        6.626          Durbin-Watson:                   2.261
Prob(Omnibus):                  0.036          Jarque-Bera (JB):                5.666
Skew:                           0.768          Prob(JB):                       0.0588
Kurtosis:                       3.503          Cond. No.                         28.3
=====================================================================================
\end{verbatim}
\end{labelverb}
}
\\
Output \ref{verb:olsreg1}. Multivariate linear Regression (Ordinary Least Squares) between the polyphenols mean and the values Texture contrast, Texture homogeneity, Texture energy, and Mean color channel 2 \\

In Output \ref{verb:olsreg2} a multivariate linear regression (Ordinary Least Squares) shows a correlation also between the mean value of FRAP mean and the values Mean color channel 2, Mean color channel 1 and Texture energy, supported by the Actual vs Predicted values image in Figure \ref{fig:multivar2} and by the related Q-Q plot of the residuals in Figure \ref{fig:qq2}  again showing the alignment along the diagonal line, indicating that the distribution of the residuals closely adheres to the theoretical distribution and telling that the model is appropriate and that the errors are random and not systematic.

{\scriptsize
\begin{labelverb}\label{verb:olsreg2}
\begin{verbatim}
OLS Regression Results                            
==============================================================================
Dep. Variable:              frap_mean   R-squared:                       0.572
Model:                            OLS   Adj. R-squared:                  0.545
Method:                 Least Squares   F-statistic:                     21.40
Date:                Sun, 09 Mar 2025   Prob (F-statistic):           6.10e-09
Time:                        16:13:18   Log-Likelihood:                 21.221
No. Observations:                  52   AIC:                            -34.44
Df Residuals:                      48   BIC:                            -26.64
Df Model:                           3                                         
Covariance Type:            nonrobust                                         
==============================================================================
                          coef    std err      t      P>|t|   [0.025    0.975]
------------------------------------------------------------------------------
const                    0.2568    0.023    11.059    0.000    0.210    0.304
Mean color channel 2     0.4291    0.104     4.135    0.000    0.220    0.638
Mean color channel 1    -0.2851    0.104    -2.745    0.008   -0.494   -0.076
Texture energy           0.0866    0.023     3.714    0.001    0.040    0.133
==============================================================================
Omnibus:                        0.157   Durbin-Watson:                   2.461
Prob(Omnibus):                  0.924   Jarque-Bera (JB):                0.305
Skew:                           0.112   Prob(JB):                        0.859
Kurtosis:                       2.699   Cond. No.                         8.84
==============================================================================
\end{verbatim}
Output \ref{verb:olsreg2}. Multivariate linear Regression (Ordinary Least Squares) between the polyphenols mean and the values Texture contrast, Texture homogeneity, Texture energy, and Mean color channel 2 \\
\end{labelverb}
}

\begin{figure}[ht!]
\centering
\includegraphics[width=0.8\textwidth]{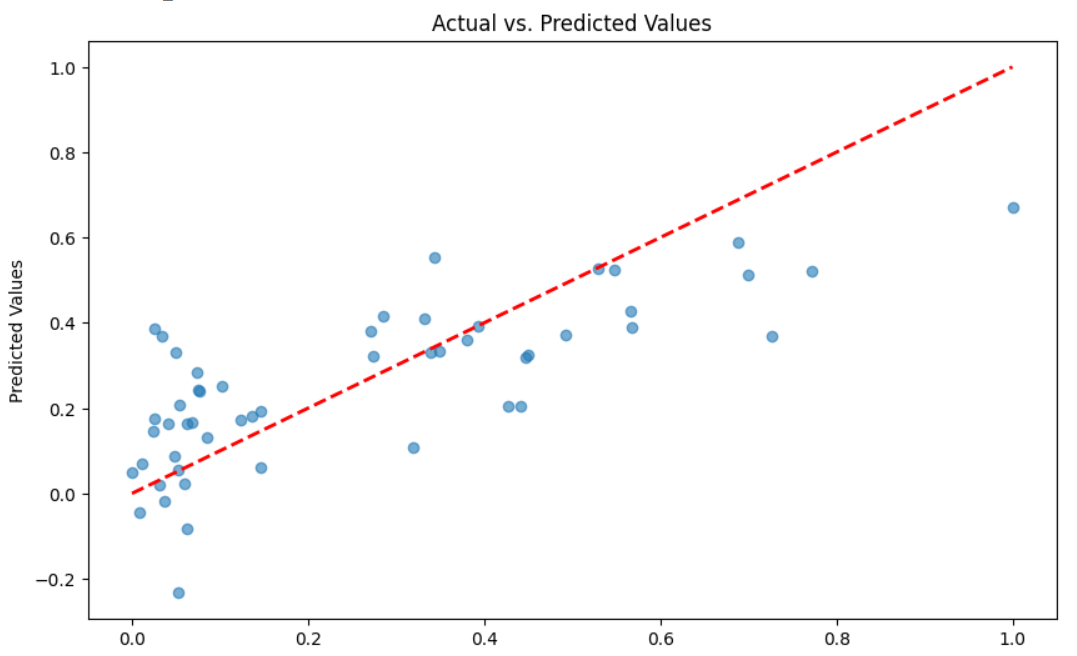}
\caption{Multivariate linear Regression (Ordinary Least Squares) between the FRAP mean and the values Texture contrast, Texture homogeneity, Texture energy, and Mean color channel 2}\label{fig10}
\end{figure}

\begin{figure}[ht!]
\centering
\includegraphics[width=0.8\textwidth]{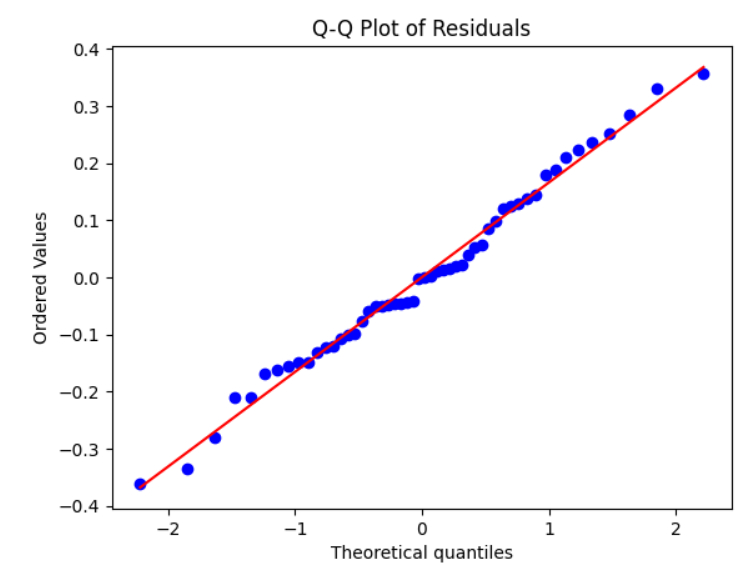}
\caption{Multivariate linear Regression plot of actual versus predicted values (Ordinary Least Squares) between the polyphenols mean and the values Texture contrast, Texture homogeneity, Texture energy, and Mean color channel 2}\label{fig:qq2}
\end{figure}

Figure \ref{fig:qq2}. Q-Q plot of the residuals of the linear Regression plot of actual versus predicted values (Ordinary Least Squares) between the polyphenols mean and the values Texture contrast, Texture homogeneity, Texture energy, and Mean color channel 2

A significant classification was also obtained using XGBoost, comparing features extracted from RGB images with fatty acids 14:0iso, C22:0, C16:0, C18:1t19, C182t12, as shown in Figs. 10 and 11, demonstrating that the model can effectively distinguish between the different categories analysed.

\begin{figure}[ht!]
\centering
\begin{minipage}{0.31\textwidth}
\centering
\includegraphics[width=\textwidth]{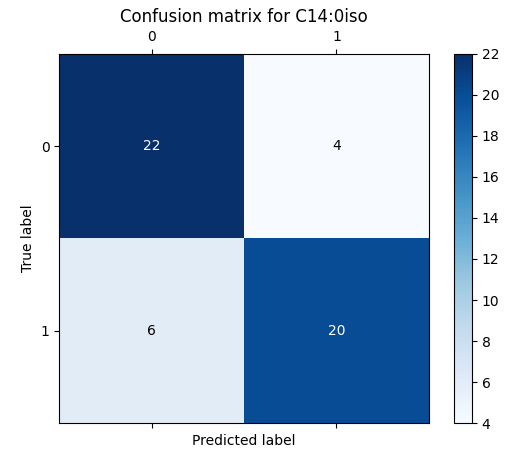}
\label{fig10a}
\end{minipage}\hfill
\begin{minipage}{0.31\textwidth}
\centering
\includegraphics[width=\textwidth]{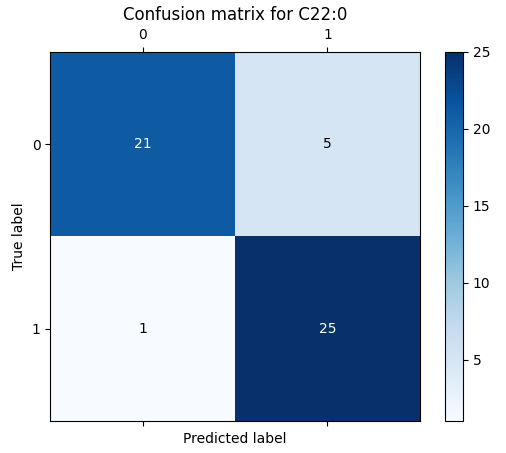}
\label{fig10b}
\end{minipage}
\begin{minipage}{0.31\textwidth}
\centering
\includegraphics[width=\textwidth]{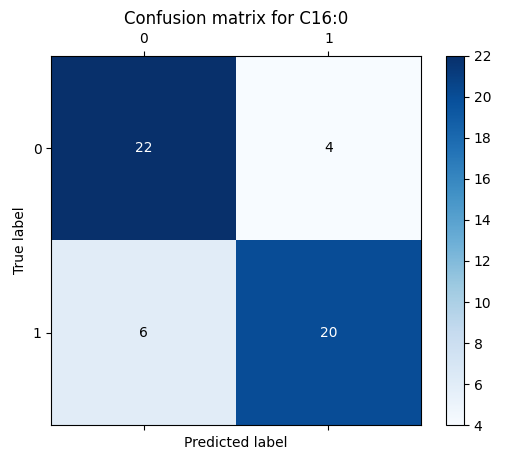}
\label{fig10c}
\end{minipage}
\caption{Confusion matrix for image classification vs C14:0iso, C22:0, C16:0}
\label{fig10}
\end{figure}


\begin{figure}[ht!]
\centering
\begin{minipage}{0.48\textwidth}
\centering
\includegraphics[width=\textwidth]{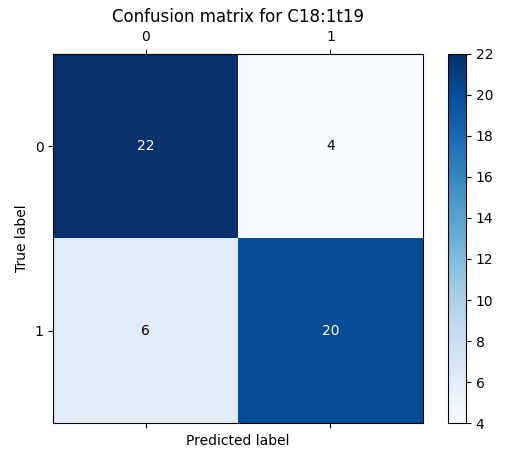}
\label{fig11a}
\end{minipage}\hfill
\begin{minipage}{0.48\textwidth}
\centering
\includegraphics[width=\textwidth]{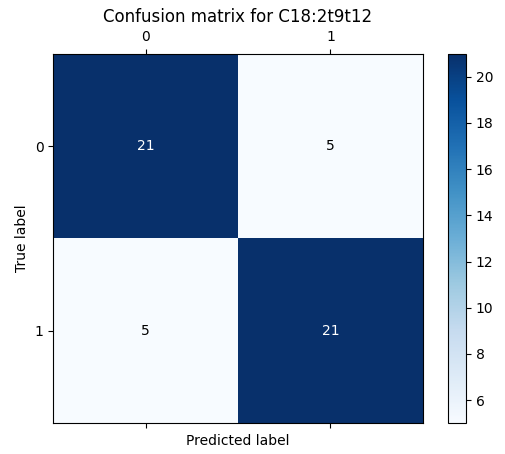}
\label{fig11b}
\end{minipage}
\caption{Confusion matrix for image classification of C18:1t19, C182t12}
\label{fig10}
\end{figure}

\subsection{Hyperspectral analysis}
Hyperspectral data from 49 samples (3 were excluded due to an alteration) enabled us to identify several measures using different Machine learning methods. Table \ref{tab:hyperallmethods} summarizes the significance results obtained by the methods utilised.

\begin{sidewaystable}
{\tiny
\centering
\begin{tabular}{lcccccccccccccc}
\toprule
Measure & Single & Sing. & PCA & Grid & SVM & MLP & Random & k-NN & PLS & LSTM & RNN & LASSO & Cluster \\
& bands & bands &  & search &  &  & 1DCNN & ANOVA \\
& significance & (Tau) &  & SVM &  &  &  &   &  &  &  & XGB & p-val \\
& Pearson &  &  &  &  &  &  &   &  &  &  &  &   \\
\midrule
polyphenols & P-val<0.05 & P-val<0.05 & & & & & Acc.>95\% & & Acc.>95\% & Acc.>95\% & Acc.>95\% & & P-val<0.05 \\
FRAP & & P-val<0.05 & & & Acc.>95\% & & & & Acc.>95\% & Acc.>95\% & Acc.>95\% & Acc.>95\% & P-val<0.05 \\
C4:0 & & & P-val<0.05 & & Acc.>95\% & & & & & & & & \\
C6:0 & P-val<0.05 & P-val<0.05 & P-val<0.05 & & Acc.>95\% & Acc.>95\% & Acc.>95\% & & & & & & \\
C8:0 & P-val<0.05 & P-val<0.05 & & Acc.>95\% & Acc.>95\% & & & & & & & & P-val<0.05 \\
C10:1c9 & P-val<0.05 & P-val<0.05 & P-val<0.05 & Acc.>95\% & Acc.>95\% & & & & Acc.>95\% & & & & P-val<0.05 \\
C13:0ante & P-val<0.05 & P-val<0.05 & P-val<0.05 & & & & & & Acc.>95\% & Acc.>95\% & Acc.>95\% & & P-val<0.05 \\
C13:0 & & & P-val<0.05 & Acc.>95\% & Acc.>95\% & & & Acc.>95\% & & & & & \\
C14:0iso & P-val<0.05 & P-val<0.05 & P-val<0.05 & & & & & & & & & & \\
C14:0 & P-val<0.05 & P-val<0.05 & P-val<0.05 & & & & & & & & & & \\
C15:0iso & P-val<0.05 & P-val<0.05 & P-val<0.05 & & & & & & Acc.>95\% & Acc.>95\% & Acc.>95\% & & P-val<0.05 \\
C14:1c9 & P-val<0.05 & P-val<0.05 & P-val<0.05 & Acc.>95\% & Acc.>95\% & & & & Acc.>95\% & & & & P-val<0.05 \\
C15:0 & & & & Acc.>95\% & Acc.>95\% & & & Acc.>95\% & & & & & & \\
C16:0iso & P-val<0.05 & P-val<0.05 & P-val<0.05 & & & & & & Acc.>95\% & & & & P-val<0.05 \\
C16:0 & P-val<0.05 & P-val<0.05 & P-val<0.05 & & & & & & & & Acc.>95\% & & P-val<0.05 \\
C17:0 & P-val<0.05 & P-val<0.05 & P-val<0.05 & Acc.>95\% & Acc.>95\% & Acc.>95\% & Acc.>95\% & Acc.>95\% & & & & & P-val<0.05 \\
C18:0 & P-val<0.05 & P-val<0.05 & P-val<0.05 & & Acc.>95\% & & & & Acc.>95\% & & & & & \\
C18:1t10 & & & P-val<0.05 & Acc.>95\% & Acc.>95\% & & Acc.>95\% & Acc.>95\% & & & & & & \\
C18:1t16 & P-val<0.05 & P-val<0.05 & P-val<0.05 & & & & & & Acc.>95\% & & & & P-val<0.05 \\
C18:3n3 & P-val<0.05 & P-val<0.05 & P-val<0.05 & & & & & & Acc.>95\% & & & & & \\
\bottomrule
\hline
\end{tabular}
\caption{Statistically significant methods using hyperspectral data}
\label{tab:hyperallmethods}
}
\end{sidewaystable}

Determining the above properties is fundamental to understanding the biological and nutritional content, which is essential for their analysis using AI or ML.
As introduced in the Methods section, various methods were employed to assess the correlation and classification of the main parameters of the milk, including the significance of single bands using Pearson and Tau correlation coefficients, Principal Component Analysis (PCA), Grid Search combined with Support Vector Machine (SVM), Support Vector Machine for classification, Multilayer Perceptron (MLP), Random Forest, k-Nearest Neighbors (k-NN), Partial Least Squares (PLS), Long Short-Term Memory (LSTM), Recurrent Neural Network (RNN), and a hybrid approach combining Least Absolute Shrinkage and Selection Operator (LASSO) with One-Dimensional Convolutional Neural Network and XGBoost (LASSO\_1DCNN-XGB). Additionally, cluster analysis was performed, with the significance of clusters evaluated using ANOVA p-values. These comprehensive methods allowed for an in-depth evaluation of the relationships and classification of the key parameters of milk.

Moreover, in a first round of Random Forest classification to distinguish between CTR, ASIG, and SIG groups and the hyperspectral curves, we achieved 100\% accuracy. We run a second round, strongly reducing the number of trees and leaves to avoid overfitting, and the results, similar to the ones obtained using the visible RGB images, are shown in Figure 13:

\begin{table}[htbp!]
\centering
\caption{Classification Report Random Forest (Accuracy: 94.83\%). Confusion matrix for hyperspectral images of the ASIG, CTR, and SIG cow groups}
\label{tab:rf-classification-report}
\begin{tabular}{lcccc}
\toprule
\textbf{GROUP} & \textbf{Precision} & \textbf{Recall} & \textbf{F1-score} & \textbf{Support} \\
\midrule
ASIG            & 0.93               & 0.87            & 0.90              & 13               \\
CTR             & 1.00               & 0.95            & 0.98              & 18               \\
SIG             & 0.92               & 1.00            & 0.96              & 18               \\
\midrule
\textbf{Accuracy} &                    &                 & \textbf{0.95}     & \textbf{49}      \\
\textbf{Macro avg} & \textbf{0.95}     & \textbf{0.94}   & \textbf{0.94}     & 49               \\
\textbf{Weighted avg} & \textbf{0.95} & \textbf{0.95}   & \textbf{0.95}     & 49               \\
\bottomrule
\end{tabular}
\end{table}

\begin{table}[ht!]
\centering
\caption{Random Forest Evaluation for hyperspectral images for GROUPS}
\label{tab:rf_eval}
\begin{tabular}{|p{2.5cm}|c|c|c|c|}
\hline
\textbf{Parameter} & \textbf{Precision} & \textbf{Recall} & \textbf{F1-score} & \textbf{Support} \\
\hline
ASIG & 0.93 & 0.87 & 0.90 & 13 \\
\hline
CTR & 1.00 & 0.95 & 0.98 & 18 \\
\hline
SIG & 0.92 & 1.00 & 0.96 & 18 \\
\hline
\multicolumn{2}{|r|}{\textbf{accuracy}} & \textbf{0.95} & & 49 \\
\hline
\multicolumn{2}{|r|}{\textbf{macro avg}} & & \textbf{0.94} & 49 \\
\hline
\multicolumn{2}{|r|}{\textbf{weighted avg}} & & \textbf{0.95} & 49 \\
\hline
\end{tabular}
\end{table}


\vspace{1em}






\section{Discussion}
The longitudinal analysis indicates that time is the primary determinant of variation in antioxidant parameters, with consistent declines in polyphenol and FRAP values across all cow groups. 
This effect of time on parameters is likely due to a physiological adaptation to the milking process, rather than a specific response to treatment.

Although treatments had a negligible impact on mean antioxidant levels, a significant result was observed when considering biological noise and system stability. The standard deviation and coefficient of variation of the FRAP values showed significant effects of the treatment and the two milking times: the mean values did not change, but they were less dispersed, suggesting that the treatments improved the animals' capacity to respond more robustly and coordinately to stress
Multivariate analysis further confirms this hypothesis, showing substantial overlap among the cow groups. Therefore, this suggests that the treatments influence the system's stability rather than its total antioxidant capacity. These results imply that the treatments improve the predictability of physiological status, reducing the range of individual variability.

From a nutritional point of view, the intake of fatty acids from milk can help reduce the risk of certain metabolic and cardiovascular diseases, thanks to the presence of medium- and short-chain saturated fatty acids. However, excessive consumption can increase saturated fat and calorie intake, promoting the accumulation of low-density lipoprotein cholesterol and the risk of heart disease.
This is why it is important to determine their presence for product quality assessment.
Most fatty acids in milk are composed of a chain of an even number of carbon atoms, ranging from 4 to 28. The predominant form is triglycerides, followed by phospholipids and cholesterol esters, which serve as important sources of energy in the diet and as essential structural elements of cells.
Short and medium-chain fats C4:0, C6:0, C8:0 and 10:1c9 are produced by the cow during digestion, and their quantity, in addition to indicating the animal's health, determines the quality of its diet. For example, butyric acid C4:0 is known for its benefits to the gut, while medium-chain fats help keep microbes at bay and strengthen the immune system.
Long-chain saturated fats such as C14:0, C14:0iso, C16:0, and C16:0iso, which derive partly from the cow's diet and partly from its reserves, are important for the consistency of milk but are also primarily responsible for those indices that measure the heart health risk of those who drink it. The exception is stearic acid C18:0, which has no negative effects on cholesterol.
Fats with odd or branched chains, such as C13:0, C13:0ante, C15:0, C15:0iso, C17:0, and C21:0, are considered the true signature of the bacteria living in the rumen, and their increased presence is a clear sign that the animal has eaten a lot of forage.
Among the unsaturated fatty acids, C14:1c9, C18:1t10, and C18:1t1,6, and in particular conjugated linoleic acid in its most active form, C18:2c9t11, are known for their potential protective effects. Omega-3 fats, such as C18:3n3 and C20:5n3, are essential for our health and increase significantly when the cow has eaten grass. Polyphenols also contribute to milk's beneficial properties thanks to their well-known antioxidant and anti-inflammatory effects. They are also powerful traceability indicators, as they are present in milk when the animal has grazed on fresh grass and hay. Finally, the Ferric Reducing Antioxidant Power is a measure that indicates the total antioxidant capacity of milk given by the presence of polyphenols, vitamins C and E and various proteins, which in addition to providing a healthy contribution to the diet, serve to prolong the shelf life of milk by protecting the properties contained in it and which help protect those who drink it from oxidative stress. Therefore, fatty acids, polyphenols, and FRAP analysis provide a “control panel" indicating the diet and health status of the animal, the quality of the milk, and, consequently, its impact on human health. 
To measure the properties of this "control panel", the following are the methods generally used, each with its own advantages and disadvantages:
\begin{itemize}
\item Gas chromatography is highly precise for quantifying fatty acids in milk, but the instrumentation is expensive, requires specialized personnel, and, above all, is time-consuming, as it often requires modifying the molecules (derivatization) to analise the sample;
\item Raman spectroscopy allows for getting simultaneous information on multiple components; however, in some cases, it can damage the sample, and the Raman signal can easily be masked by the sample's fluorescence;
\item Classical methods, such as chemical and enzymatic methods for parameters such as lactose and protein, are simple, rapid, inexpensive, widely accepted, and often used as official references. They offer high specificity in some cases, but are limited in their applicability to a single parameter. They can also be less sensitive than advanced techniques and are subject to interference, which reduces their accuracy.
\end{itemize}

The results of our study not only confirm the effectiveness of hyperspectral analysis for analyzing, correlating, and detecting the presence or absence of chemical compounds in milk, but also introduce a cutting-edge, non-destructive integration system combining visible-range hyperspectral imaging with Artificial Intelligence methods to assess its quality. 
The application of a wide range of machine learning models, such as PCA, SVM, MLP, Random Forest, and the hybrid LASSO\_1DCNN-XGB, not only enabled quantification but, more importantly, provided interpretive understanding of the spectral signatures.
The high Pearson correlation coefficients observed for many fatty acids indicate a linear relationship between their concentrations and the spectral signal, compared with classical quantitative methods; at the same time, the significance of Tau correlation coefficients indicates non-linear relationships, justifying the use of advanced ML/AI models in these cases. Using rigorous optimization to mitigate the risk of overfitting, the Random Forest model provided a remarkable accuracy of 94.83\% in distinguishing between the CTR, ASIG, and SIG groups; several ML/AI models achieved in many cases an accuracy above 95\% or even the remarkable 100\% classification accuracy in the case of two fatty acids (Figure 13); some spectral biomarkers were also isolated for nutritionally relevant fatty acids (e.g., C18:1t19 and C18:2t12).
A distinctive novelty of our approach is the effectiveness achieved with only visible images, which enabled samples to be distinguished by shelf life, antibiotic presence, and polyphenol quantity estimation. 
In a sector where the quick and accurate identification of critical compounds and where adulteration detection is fundamental, this non-destructive, multi-analytical, agile, economical and accurate quality control approach also demonstrates an enormous potential of the proposed methods, allowing an instant and clear assessment of milk quality, overcoming the time and cost limitations of traditional analytical systems,  suggesting the possibility of developing portable, low-cost devices for first-level controls, integrated with more detailed hyperspectral analysis.
To consolidate the scientific validity and applicability on an industrial scale, it is essential that future analyses use larger sample sizes or employ data augmentation and cross-validation techniques to improve the generalisability of the models. Additionally, we intend to expand these analyses to include products derived from milk.

\section{Conclusions}

Visible and hyperspectral data open new frontiers in remote sensing and material analysis by providing rich spectral information beyond human vision. Detailed spectral information enables precise identification and characterization of materials across various fields, particularly in livestock farming.
The benefits of the applied techniques are clear: these methods are non-destructive, meaning the milk sample remains unaltered and requires no preparation. They deliver rapid, real-time results, making them ideal for seamless integration into quality control processes throughout the production chain, from milking to the finished product. Additionally, they are multi-analytical, capable of simultaneously determining multiple parameters from a single scan, providing a comprehensive overview of the composition and distribution of components.
In the long run, this translates into significant cost savings compared to traditional laboratory analyses, along with a substantial enhancement in food safety.
Of course, like any cutting-edge technology, hyperspectral imaging has its challenges. The amount of data generated is immense, requiring sophisticated algorithms to process it. Extracting meaningful information requires specific expertise in artificial intelligence. Initial equipment costs can be high, although they are gradually decreasing. Creating robust predictive models requires large and diverse calibration datasets, and there is still no international standardization of validation protocols. Furthermore, near-infrared light has limited penetration into thick liquids such as milk, which can impact accuracy for very thick samples.
Despite the challenges, the field is seeing rapid advances. Integration with artificial intelligence is revolutionizing the accuracy and efficiency of hyperspectral data analysis, enabling the identification of complex patterns and automating decision-making. We are seeing the development of portable and miniature devices, making this technology more accessible and suitable for field use or on smaller production sites. Data fusion techniques are being explored, combining hyperspectral with other methodologies to obtain even more comprehensive information.
Ultimately, hyperspectral imaging is an exciting frontier for milk analysis that is maturing at an impressive rate, driven by advances in artificial intelligence. Its potential to improve quality control, ensure food safety, and optimize operational efficiency across the entire dairy supply chain is immense and will continue to expand in the coming years. It is fascinating to think about how technology allows us to "see" what is invisible to the naked eye, making the milk we consume every day increasingly safe and of guaranteed quality.
In our study, we demonstrated the usefulness of visible analysis that allowed us to characterize the presence or absence of compounds with an accuracy of 100\% or slightly less using explainable machine learning techniques.
In the dataset used for hyperspectral analysis, the altered condition of the received samples, along with the limited sample size, constrained the robustness and broader applicability of the findings. To strengthen the reliability and scientific validity of the results, further investigations with a substantially larger and properly controlled sample set are essential.


\appendix
\section{Supplementary Material}
\label{app1}

Supplementary material text.

\begin{table}[ht!]
\caption{ LMM Linear Mixed Model }
\centering
\noindent
\makebox[\textwidth][c]{
\begin{tabular}{@{}lcccccccccccccc@{}}
\toprule
Parameter & TIME\_p  & GROUP\_p & INT\_p \\
\midrule
frap & 1.632230e-09 & 0.101937 & 0.229765 \\
polyphenols & 2.890032e-08 & 0.338167 & 0.466654 \\
FA\_POLY & 2.263319e-01 & 0.537033 & 0.263347 \\
OMEGA6 & 2.406110e-01 & 0.717274 & 0.306702 \\
OMEGA3 & 2.783735e-01 & 0.152915 & 0.205767 \\
FA\_SAT & 4.727092e-01 & 0.893283 & 0.882075 \\
FA\_MONO & 6.114462e-01 & 0.714230 & 0.580648 \\
\bottomrule
\end{tabular}
}
\end{table}


\bibliographystyle{elsarticle-harv} 
\bibliography{biblio.bib}




\end{document}